\documentclass[a4paper,fleqn,usenatbib]{mnras}

\usepackage{newtxtext,newtxmath}

\usepackage[T1]{fontenc}
\usepackage{ae,aecompl}

\usepackage{graphicx}   
\usepackage{amsmath}    
\usepackage{tikz,xcolor,hyperref}
\usepackage{multirow}
\usepackage{makecell}

\usepackage{pdflscape}

\definecolor{lime}{HTML}{A6CE39}
\DeclareRobustCommand{\orcidicon}{%
    \begin{tikzpicture}
    \draw[lime, fill=lime] (0,0)
    circle [radius=0.16]
    node[white] {{\fontfamily{qag}\selectfont \tiny ID}};
    \draw[white, fill=white] (-0.0625,0.095)
    circle [radius=0.007];
    \end{tikzpicture}
    \hspace{-2mm}
}

\newcommand{\orcidAM}{\href{https://orcid.org/0000-0002-9382-2542}{\orcidicon}}
\newcommand{\orcidWS}{\href{https://orcid.org/0000-0002-2393-8427}{\orcidicon}}
\newcommand{\orcidJDD}{\href{https://orcid.org/0000-0001-9704-6408}{\orcidicon}}
\newcommand{\orcidPKS}{\href{https://orcid.org/0000-0003-2244-1512}{\orcidicon}}

\newcommand{\kms}{{\mathrm{km~s^{-1}}}}

\newcommand{\Msun}{M$_{\odot}$} 
\newcommand{\Rsun}{R$_{\odot}$} 
\newcommand{\cpd}{d$^{-1}$} 
\newcommand{\dsct}{$\delta$\,Sct}

\newcommand{\MESAbinary}{\texttt{MESA-binary}}
\newcommand{\PHOEBE}{\texttt{PHOEBE\,2}}

\newcommand\T{\rule{0pt}{2.6ex}}       
\newcommand\B{\rule[-1.2ex]{0pt}{0pt}} 

\usepackage{color}
\definecolor{todo}{rgb}{0.89,0.0,0.13}

\usepackage{color}
\definecolor{green(new)}{RGB}{50, 200, 60}

\usepackage{color}
\definecolor{cadetgrey}{rgb}{0.57, 0.64, 0.69}

\usepackage{color}
\definecolor{oror}{RGB}{0,150,0}

\newcommand{\comment}[1]{\iffalse{#1}\fi}

\title[AB\,Cas]{The eclipsing binary systems with $\delta$\,Scuti component -- II. AB\,Cas}

\author[A. Miszuda et al.]{
A. Miszuda\thanks{E-mail: miszuda@astro.uni.wroc.pl (AM)}\orcidAM,
P. A. Ko\l aczek-Szyma\'nski\orcidPKS,
W. Szewczuk\orcidWS,
J. Daszy\'nska-Daszkiewicz\orcidJDD
\\
University of Wroc{\l}aw, Faculty of Physics and Astronomy, Astronomical Institute, ul. Kopernika 11, 51-622 Wroc{\l}aw, Poland\\
}

\date{Accepted XXX. Received YYY; in original form ZZZ}

\pubyear{2021}

\begin{document}
\label{firstpage}
\pagerange{\pageref{firstpage}--\pageref{lastpage}}
\maketitle

\begin{abstract}
We present a complex study of the eclipsing binary system, AB\,Cas. The analysis of the whole \textit{TESS} light curve, corrected for the binary effects, reveals 112 significant frequency peaks with 17 independent signals. The dominant frequency $f_1 = 17.1564$\,\cpd\ is a radial fundamental mode. The $O-C$ analysis of the times of light minima from over 92 years leads to a conclusion that due to the ongoing mass transfer the system exhibits a change of the orbital period at a rate of 0.03\,s per year.
In order to find evolutionary models describing the current stage of AB\,Cas, we perform binary evolution computations. Our results show the AB\,Cas system as a product of the rapid non-conservative mass transfer with about 5-26\% of transferred mass lost from the system. This process heavily affected the orbital characteristics of this binary and its components in the past. 
In fact, this system closely resemble the formation scenarios of EL\,CVn type binaries.
For the first time, we demonstrate the effect of binary evolution on radial pulsations and determine the lines of constant frequency on the HR diagram. From the binary and seismic modelling, we obtain constraints on various parameters. In particular, we constrain the overshooting parameter, $f_{\rm ov}\in [0.010,~0.018]$, the mixing-length parameter, $\alpha_{\rm MLT}\in[1.2,~1.5]$ and the age, $t \in [2.3,~3.4]$\,Gyr.

\end{abstract}




\section{Introduction}\label{sec:introduction}

Multiple and binary systems make up a vast majority of all observed medium- and high-mass stellar objects in our Galaxy \citep{Duchene2013}. However, it is safe to say that generally most of stellar objects ranging in various masses reside in multiple systems.
Double-lined eclipsing binaries (DLEBs) are a particular part of this group of objects. They happen to be one of the most supreme astrophysical tools as they provide a unique opportunity to determine masses and radii of the components with an outstanding accuracy, often with about 1\% of errors \citep[see e.g.][]{Torres2010}. With such accuracy, DLEBs are used as benchmarks for testing the theory of stellar evolution, in particular for the age determination \citep[see e.g.][]{Higl2017,Daszynska2019}. Of special interest are the binary systems with pulsating components as they can provide independent constraints on parameters of a model and theory.

The computations of the binary evolution date back to the late 1950s, when the early binary-evolution theory was published by \cite{Kopal1959}. The theory was followed by the first binary-evolution models calculated by \cite{Morton1960}, who also confirmed the \textit{Algol paradox} theory of \cite{Crawford1955}. The first techniques for calculating the rate of mass loss during the mass transfer (MT) from the star filling its Roche lobe (donor) were described by \cite{Paczynski1970,Paczynski1971}. Many contributions on binaries were also presented at the Trieste Colloquium on "Mass Loss from Stars" \citep{Hack1969} and at the IAU Colloquium No.\,6 on "Mass Loss and Evolution in Close Binaries" \citep{Gyldenkerne1970}. Until now, there are plenty of available codes calculating the binary evolution, e.g., \cite{Zahn1977}, \cite{Wellstein2001} and \cite{Hurley2002}, with the most recent one, the \MESAbinary\ code of \cite{Paxton2015}.

Binary interactions that occur in close systems can have a significant impact on the binary's structure and evolution. One of the most prominent large-scale effect is a mass transfer between the components which effectively changes the mass distribution in close binaries. While stripping the donor of its outer layers and accreting a fresh material on the acceptor it "rejuvenates" the initially less-massive star. Although, the scientific discussion still continues, whether MT should be considered as a fully conservative process \citep[eg.][]{Kolb1990,Sarna1992,Sarna1993,Guo2017}, it has been shown by \cite{Chen2017} that the formation of systems similar to AB\,Cas can be satisfactorily explained with non-conservative MT, with about 50\% of transferred mass lost from the system. On the other hand, recent work of \cite{Miszuda2021} explained the formation and evolution of KIC\,10661783, a system of EL\,CVn type containing a \dsct\ primary, as an effect of nearly-conservative mass transfer, with only 5\% of the transferred mass lost from the system.

The $\delta$\,Scuti (\dsct) stars are the pulsators of intermediate-mass ($1.5 - 2.5\,\rm M_{\odot}$) located within the classical instability strip. They are mainly in the main sequence phase
of evolution, however there are known cases of \dsct\ stars evolving through the Hertzsprung gap and in the pre-main sequence phase \citep[e.g.][]{Rodriguez2000,Dupret2005,Aerts2010,Liakos2017,Murphy2018,Murphy2019}.
The majority of \dsct\ stars pulsate in low-order pressure (p) modes excited by the $\kappa$ mechanism, with periods shorter than 0.3\,d. 
That makes them easily recognisable from the nearby located in the HR diagram $\gamma$\,Doradus stars, pulsating in high-order gravity (g) modes. 
The region of the $\gamma$\,Doradus pulsators partially overlaps with the $\delta$\,Scuti instability strip in the HR diagram, resulting in \dsct/$\gamma$\,Dor hybrids \citep[e.g.][]{Grigahcene2010,Balona2015,Antoci2019} pulsating in both p and g modes simultaneously.

The \dsct\ pulsators in semi-detached binary systems actually form a separate subclass, the so-called oEA (oscillating Eclipsing Algols) stars \citep{Mkrtichian2004}. 
The evolution of oEA stars significantly differs from the evolution of the classical \dsct\ stars in detached binaries.
\cite{Mkrtichian2004} suggested that a rapid mass transfer and accretion by the pulsating gainer can potentially change its oscillation properties. This hypothesis has been recently confirmed by \cite{Miszuda2021}, who showed that the binary evolution can lead to the helium enrichment in the outer layers of the acceptor. This in turn has a huge impact on the frequencies and the excitation of the pulsation modes.

AB\,Cas \citep[A3V + K1V][]{Rodriguez2001} is an Algol-type eclipsing variable star discovered by \cite{Hoffmeister1928}. The system was observed during many photometric campaigns, e.g., by \cite{Tempesti1971}, \cite{Soydugan2003}, \cite{Rodriguez2004data} and \cite{Abedi2007}. It was found by \cite{Tempesti1971} that the main component undergoes the periodic changes in brightness with an amplitude of 0.05\,mag in the \textit{V} filter. The period of these changes has been established on 0.0583\,d \citep{Rodriguez1998,Soydugan2003,Rodriguez2004}. The first spectroscopic observations of the system were performed by \cite{Kaitchuck1985}, who were looking for emission in hydrogen lines. Such lines could indicate the existence of the accretion disc, however, no emission lines were found in the system.
Later, \cite{Nakamura1988} determined the radial velocity amplitude of the primary component at $K_1=41.9\,\kms$.
\cite{Soydugan2003}, using the Johnson's photomety in \textit{B} filter determined that the system has a semi-detached configuration and that the mass-ratio of the components is $q\equiv M_2/M_1 \approx 0.19$. This value was later refined by \cite{Rodriguez2004}, who found the mass-ratio value at $q=0.201$. Using the spatial-filtering technique and the photometric amplitudes and phases, \cite{Rodriguez2004} identified the dominant pulsation frequency, $f=17.1564$\,\cpd, as a radial fundamental mode.
The first high-resolution spectroscopic time-series were gathered by \cite{Hong2017}. The authors analysed 27 spectra of the binary and measured the radial velocity of both components for the first time. The masses, radii and effective temperatures for the components were respectively determined at $M_1=2.01 \pm 0.02$\,\Msun, $M_2=0.37 \pm 0.02$\,\Msun; $R_1=1.84 \pm 0.02$\,\Rsun, $R_2=1.69 \pm 0.03$\,\Rsun; $T_{\rm eff,1}=8080 \pm 170$\,K and $T_{\rm eff,2}=4925 \pm 150$\,K.

This is a second paper in the series of $\delta$\,Sct stars in eclipsing binary systems. The first one \citep{Miszuda2021} was devoted to the analysis of KIC\,10661783.
Here, we present an extended study of AB\,Cas binary. In Section\,\ref{sec:observations} we give a short description of the observations that we analyse in Section\,\ref{sec:binarymodelling} in order to obtain the orbital and absolute parameters of the components. We study the change of the orbital period in Section\,\ref{sec:orb_period_variation}. Later, in Section\,\ref{sec:freqanalysis} we analyse the pulsational variability of AB\,Cas. Section\,\ref{sec:binaryevolution} is devoted to the binary-evolution modelling of the system and in Section\,\ref{sec:SeismicModelling} we perform the seismic modelling. Discussion, conclusions and future prospects in Section\,\ref{sec:conclusions} end the paper.
Finally, in Appendix\,\ref{sec:appendix} we provide a list of all significant frequencies found in the data with their amplitudes and phases.

\section{Observations}
\label{sec:observations}

In this paper we use all \textit{TESS} \citep[Transiting Exoplanet Survey Satellite,][]{TESS2009} observations that are available at the time of a paper-writing. In addition, we supplement them with the multi-colour Str\"omgren \textit{uvby} photometry obtained by \cite{Rodriguez2004data,Rodriguez2004}.

\subsection{\textit{TESS}}
\label{sec:tess-photometry}
The AB\,Cas system was, up to now, observed by the \textit{TESS} mission within a cycle 2, covering sectors 18, 19 and 25. It is planned, that \textit{TESS} will observe this system again in the near future, i.e. in late May and early June 2022, during cycle 4, in sector 52.

The 2-min \textit{TESS} data were obtained and pre-processed using the \texttt{Lightkurve}\footnote{https://docs.lightkurve.org/} code, a Python package for \textit{Kepler} and \textit{TESS} data analysis \citep{lightkurve2018}.
We extracted the flux from the Target Pixel Files (TPF) using a custom defined mask that contained all pixels showing a flux larger than three times the standard deviation above the overall median for each sector. 
Next, we corrected the light curve for the outliers using a 5$\sigma$ criterion. By an eye inspection, we rejected the points that showed obvious unphysical trends (e.g. sudden changes in brightness by several orders of flux magnitude). Then, we divided the light curve from each sector into two parts separated by observational gaps and normalized each part separately using a linear regression for the data without eclipses. At the end we merged all parts back together.

The final \textit{TESS} light curve of AB\,Cas consists of 51\,300 points for 2-min cadence, spread through nearly 218 days in sectors 18, 19 and 25. The pseudo-Nyquist frequency for these data is $f_{\rm N} \sim 360$\,\cpd\ and the Rayleigh limit is $\Delta f_{\rm R} \equiv 1/T = 0.0046$\,\cpd. We present these observations in Fig.\,\ref{lc_TESS_inset}. Additionally, in order to visualise the light variations from, both, eclipses and pulsations, in Fig.\,\ref{lc_TESS_inset} we show the inset covering the observations from sector 19 in a five days time window. The phase-folded \textit{TESS} observations are shown in Fig.\,\ref{lc_uvby} with grey points.

\subsection{Str\"omgren photometry}
\label{sec:stromgren-photometry}
\begin{figure}
    \flushright
    \includegraphics[width=\columnwidth,clip]{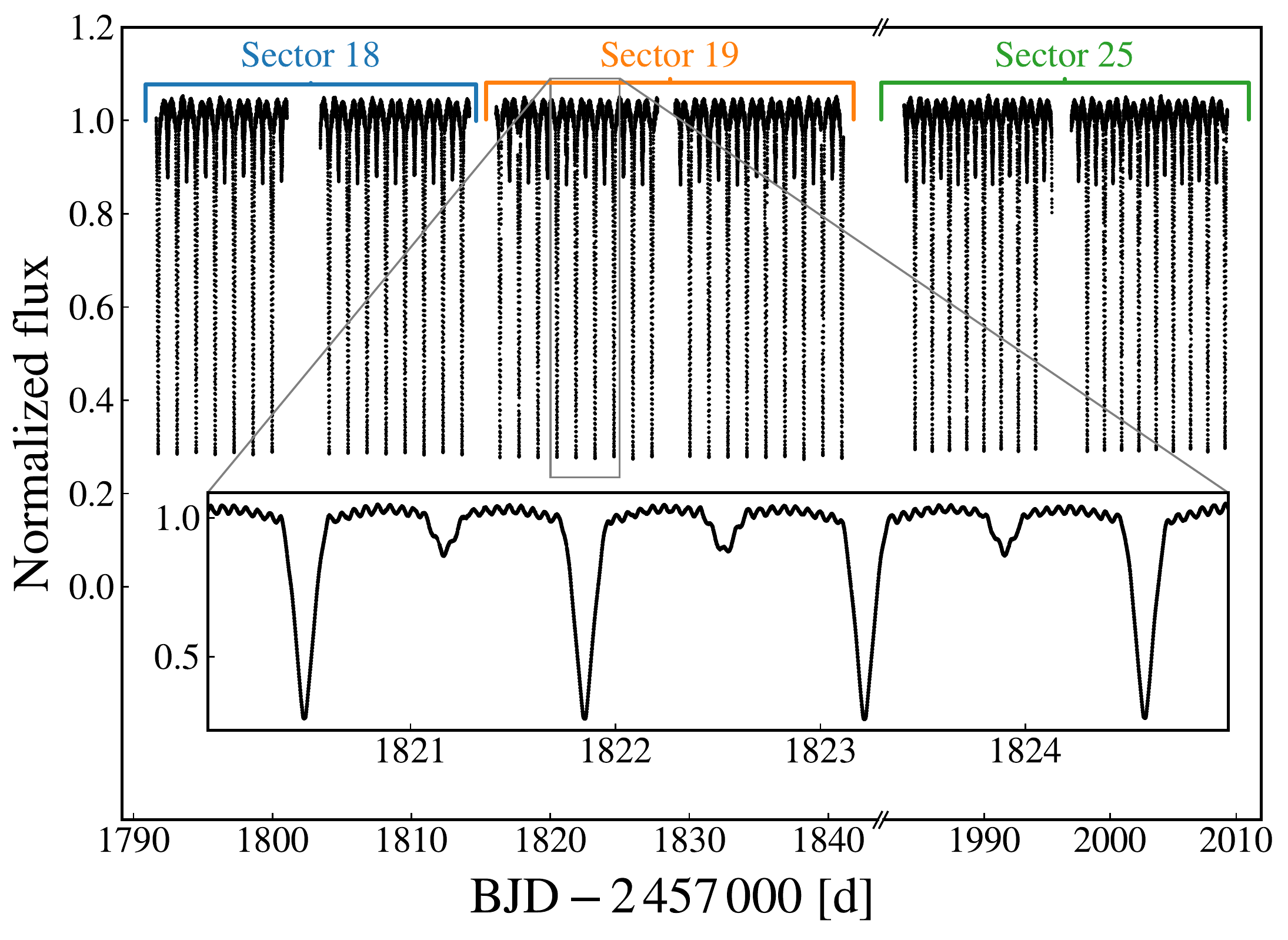}
    \caption{The full 2-min \textit{TESS} light curve of AB\,Cas from sectors 18, 19 and 25. The inset shows the zoomed area covering 5 days of observations to show both binarity and pulsational variability. }
    \label{lc_TESS_inset}
    
    \includegraphics[width=0.98\columnwidth,clip]{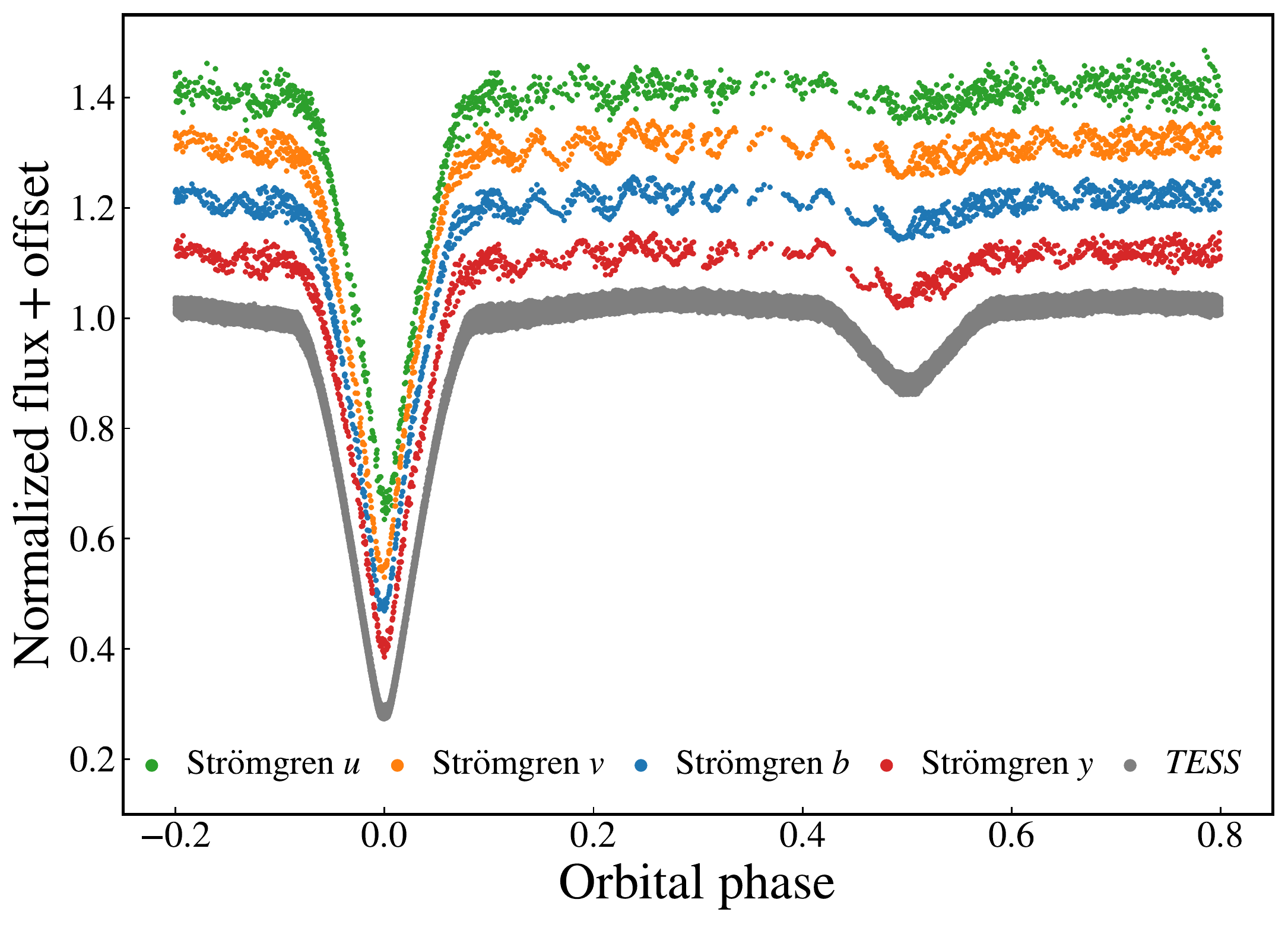}
    \caption{The phase-folded four-color Str\"omgren  and \textit{TESS} photometry. The zero-phase corresponds to the moment of superior conjunction.}
    \label{lc_uvby}
\end{figure}

In order to put independent constraints, in addition to \textit{TESS} photometry, we use the Str\"omgren four-colour photometric observations from \cite{Rodriguez2004data,Rodriguez2004}. These observations were carried out between October 1998 and November 1999 at Sierra Nevada Observatory in Spain. During 20 photometric nights, the authors collected 1313 simultaneous measurements in \textit{uvby} filters and transformed them into differential magnitudes.
Observations were performed with particular emphasis on ensuring a good coverage of the eclipses.
For further details on obtaining the data we refer to \cite{Rodriguez2004} and references therein.
We present these observations in Fig.\,\ref{lc_uvby}.
Due to the small number of points, the pulsation variability is not averaging during the phasing procedure, hence,
the internal variability is clearly visible in the systems light curve outside of the eclipses.

\section{Binary light curve modelling}
\label{sec:binarymodelling}

Until now, AB\,Cas system has been modelled solely on the basis of ground-based photometry and spectroscopy. 
However, the quality of this photometry is significantly lower when compared to the currently available \textit{TESS} space photometry. Therefore, we decided to re-fit the orbital and physical parameters of the system, taking into account the precise \textit{TESS} 2-min light curve from three sectors (Sect.~\ref{sec:tess-photometry}) and the available Str\"omgren time-series photometry (Sect.~\ref{sec:stromgren-photometry}). 
Based on our solution, presented in Table\,\ref{tab:PHOEBE_paramterers}, we perform evolutionary modelling of the system. We present this modelling in Sect.~\ref{sec:binaryevolution}.

\subsection{Preparation of the light curves}
\label{sec:binarymodelling_preparations}
First, by means of the Fourier analysis applied to the whole \textit{TESS} light curve, we determined the orbital period of AB\,Cas to be equal to $P_{\rm orb}=1.3668810(6)$\,d in the epoch of the \textit{TESS} observations. To this end, we took the orbital frequency $f_{\rm orb}$ from the periodogram  and fitted it to the data along with 50 orbital harmonics.
We adopted this value of $f_{\rm orb}$ in further modelling.

Next, we converted the \textit{uvby} magnitudes from \cite{Rodriguez2004data,Rodriguez2004} to fluxes.
Similarly as in the \textit{TESS} case (see Sect.\,\ref{sec:tess-photometry}) we normalized these data using the linear regression for the out-of-eclipse data.
Since the aforementioned Str\"omgren photometry had no measurement errors attached, we assigned each \textit{uvby} magnitude a typical ground-based photometric error of $0.01$\,mag, which is roughly an order of magnitude larger than the \textit{TESS} observational errors. 

The amplitude of the dominant pulsation mode with a frequency of about $17.15$\,d$^{-1}$ is significant when compared to the depth of both eclipses (see in Fig.~\ref{lc_TESS_inset}). Hence, we decided to subtract this distinct high-amplitude frequency from all observed light curves prior to the subsequent analysis. This is of particular importance for the \textit{uvby} light curves, which do not cover many orbital cycles, so the pulsational variability can not be averaged out satisfactorily. Consequently, when unsubtracted, it could influence the result of binary light curve modelling.

Finally, each light curve was phase-folded with the orbital period determined earlier from the \textit{TESS} data. Then, the phased light curves were binned in phase by calculating the median values of normalised fluxes in each phase bin. The bins were defined by us, however, their widths were not constant. We chose them so that the distance between the points of the binned light curve on the orbital phase -- normalized flux plane was kept approximately constant. Thanks to this procedure, narrow eclipses were sampled more densely than areas outside of them. Therefore, the out-of-eclipse variability did not dominate the solution. The binned light curves contain 150 and 94 points in the case of \textit{TESS} and each Str\"omgren-passband photometry, respectively. They are presented in Fig.~\ref{fig:phoebe-fit}a and \ref{fig:phoebe-fit}e. 

\subsection{The simultaneous fit}
In order to derive the physical and orbital parameters of AB\,Cas we performed a simultaneous fit to the five phased and binned light curves, which are described above.
For the purpose of the binary light curve modelling we used the \PHOEBE\ modelling software\footnote{http://phoebe-project.org/} \citep[PHysics Of Eclipsing BinariEs\,2, version 2.3,][]{Prsa2005,Prsa2016,Horvat2018,Jones2020,Conroy2020}. 
The fitting was realised as error-weighted least-square optimization implemented as the trust-region reflective algorithm \citep[TRF,][]{TRF}, included in the Python \texttt{Scipy} package \citep{2020SciPy-NMeth}.
We chose the TRF algorithm because it allows taking into account the boundaries of parameters' values that naturally occur in our problem.

During the fit, we assumed masses of the components after \cite{Hong2017} to be $M_1=2.01 (2)$\,M$_\odot$, $M_2=0.37(2)$\,M$_\odot$, along with the circular orbit ($e=0$), the synchronous rotation of both components, and the spin-orbit alignment in the system. The last three assumptions are justified by strong tidal interactions that we expect to act for a long time between components of close systems \citep[e.g.][]{Zahn1975,Zahn1977}.
During the preliminary modelling we noticed that when trying to model the system in a detached geometry, the equivalent radius\footnote{The radius of a sphere that has the same volume as modeled star without spherical symmetry.} $R_{\rm equiv,2}$ of the secondary always tends to the critical value at which the secondary fills its Roche lobe. 
Hence, we modelled the system in a semi-detached geometry when $R_{\rm equiv,2}$ is no more a free parameter but a function of the components' masses and the orbital period. 
We also found that solution is sensitive to the albedo $A_{\rm bol,2}$ and the gravity darkening $g_2$ of the secondary, whereas the fit was barely sensitive to $A_{\rm bol,1}$ and $g_1$. Therefore, we set $A_{\rm bol,2}$ and $g_2$ as free parameters. This allowed us to properly reconstruct the profile of the secondary eclipse. In the same time the bolometric albedo and the gravity-darkening coefficient were fixed to their typical values for the radiative envelopes, i.e.~$A_{\rm bol,1}=1$ and $g_1=1$.
Finally, the following free parameters were subjects to the optimization process: the inclination of orbit $i$, the moment of superior conjunction $T_0$, the effective temperatures of both components $T_{\rm eff,1,2}$, the equivalent radius of the primary component $R_{\rm equiv,1}$, the bolometric albedo $A_{\rm bol,2}$, and the gravity-darkening coefficient $g_2$, of the secondary component.
The surfaces of both components were simulated within \PHOEBE\ with 4000 triangular elements. Fluxes and the interpolated limb-darkening coefficients were obtained from ATLAS\,9 model atmospheres \citep{Castelli2003} with solar metallicity for the primary. In the case of secondary component, the solar-metallicity PHOENIX models \citep{phoenix1,phoenix2} were used due to the photospheric conditions of strongly-deformed surface, which were outside the range covered by the ATLAS\,9 models. Both grids of the atmosphere models are incorporated into \PHOEBE\ with the accompanying limb-darkening tables. \cite{Hong2017} reported that AB\,Cas is characterised by the colour excess, $E(B-V)=0.128$\,mag. This value has a minor, but still noticeable, impact on the shape of our model light curves, so we included effects of the interstellar extinction into our modelling, assuming a typical Galactic value of the total-to-selective extinction ratio, $R_V=3.1$. The reflection/irradiation effect is significant in the AB\,Cas system and it was treated in the formalism developed by \cite{Horvat-irradiation-effect}. Finally, using the actual transmission functions of the \textit{TESS} and Str\"omgren filters, model flux variations in individual passbands were calculated and fitted to the observational data.

\begin{figure*}
\centering
\includegraphics[width=\textwidth,clip]{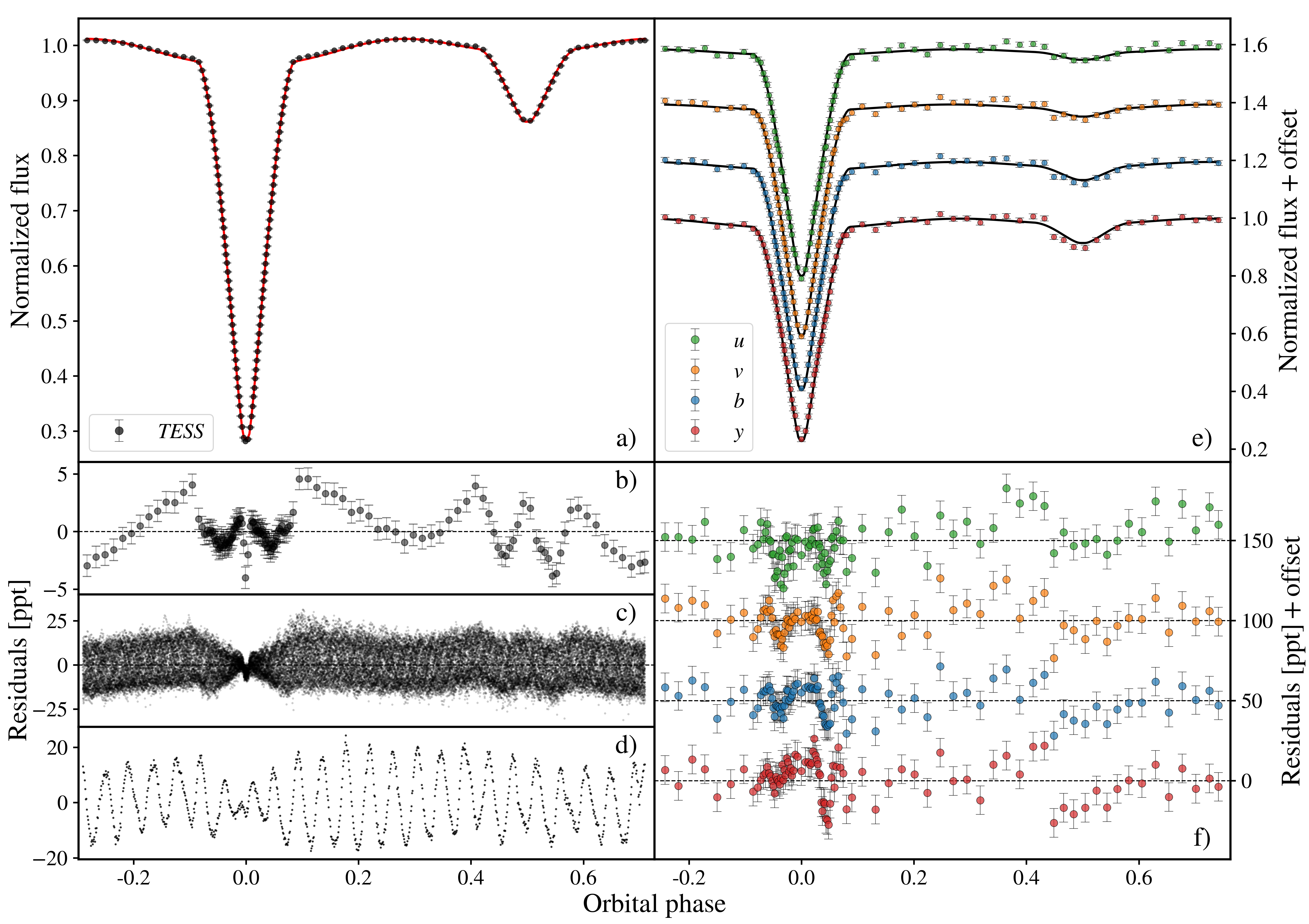}
\caption{The results of simultaneous modelling of multi-band light curves of the AB\,Cas system using the \texttt{PHOEBE\,2} software. 
Panel a): phased and binned \textit{TESS} light curve with the best-fitting model superimposed. 
Panel b): the residuals from the fit presented in panel a. 
Panel c): the same as in panel b, but for the unbinned \textit{TESS} data. 
Panel d): the same as in panel c, but for the one selected orbital cycle. 
Panel e): phased and binned \textit{uvby} light curves with the best-fitting models superimposed.
Panel f): the residuals from the fit presented in panel e. For the sake of clarity, some vertical offset was introduced in panels e and f. The zero phase on each panel corresponds to the moment of superior conjunction, i.e. $T_0 = \rm BJD\,2458791.777388$.}
\label{fig:phoebe-fit}
\end{figure*}

\begin{figure}
\centering
\includegraphics[width=\columnwidth,clip]{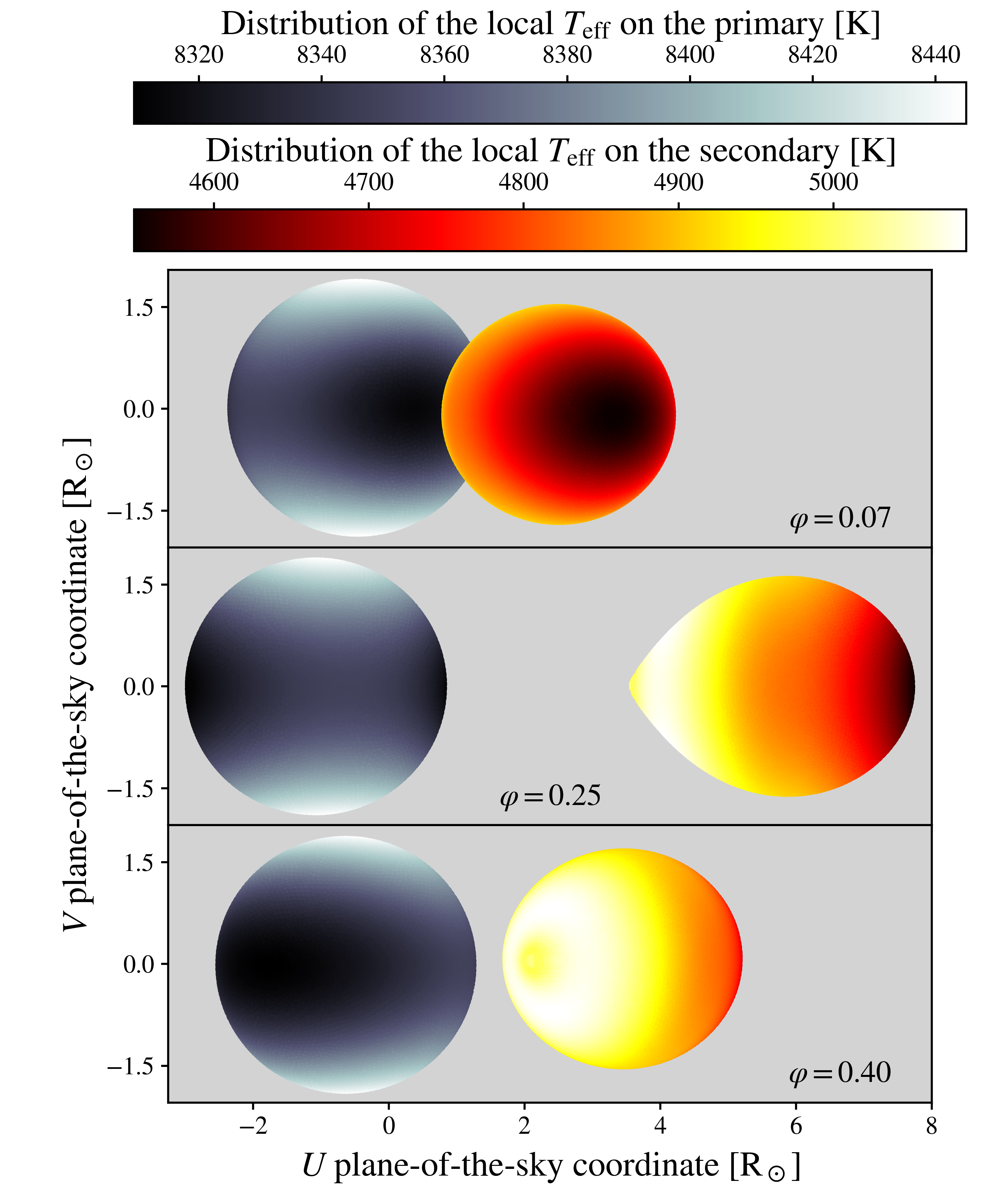}
\caption{The simulated appearance of the AB\,Cas system in the plane of the sky, corresponding to our best solution. The system is presented in three selected orbital phases, specified in each of the consecutive panels. The color-coding shows the variation of local effective temperature across the surfaces of the components. Note the separate temperature scales for the primary and secondary component which are placed at the top of the figure. Coordinates $(U,V)=(0,0)$ indicate the system's barycenter. Phase 0.0 corresponds to the moment of superior conjunction.}
\label{fig:phoebe-temperatures}
\end{figure}

The result of our fit can be seen in Fig.~\ref{fig:phoebe-fit} while the optimized values of parameters are stored in Table~\ref{tab:PHOEBE_paramterers}. 
After subtracting the best-fitting model from the \textit{TESS} data, the pulsational variability becomes clearly visible, as presented in Fig.~\ref{fig:phoebe-fit}d. One can also notice the expected reduction in the observed amplitude of the dominant pulsation mode with a frequency of $f\sim 17.15$\,d$^{-1}$ during the primary eclipse.
This amplitude reduction manifests itself as a smaller residual scattering seen in Fig.~\ref{fig:phoebe-fit}c and Fig.~\ref{fig:phoebe-fit}d. We also present the simulated appearance of the AB\,Cas system in the plane of the sky at three different orbital phases in Fig.~\ref{fig:phoebe-temperatures}.

Although the overall fit is satisfactory, especially the \textit{TESS} residual light curve (Fig.~\ref{fig:phoebe-fit}b) reveals some systematic differences between the best model and the observations of the order of 4\,ppt. Systematic deviations seem to be also present in the \textit{uvby} residuals, but since they are of poorer photometric quality when compared to \textit{TESS}, we focus only on the residuals of the latter. The remaining signal can not be explained by introducing some small, but non-zero, eccentricity. By making eccentricity a free parameter in our ancillary fits we always got values effectively equal to zero. Similarly, introducing the non-synchronous rotation, the minor spin-orbit misalignment or the third-light did not significantly improve the situation presented in Fig.~\ref{fig:phoebe-fit}b. Therefore, it seems that the systematic differences can be explained in a few ways.
Firstly, they may originate from the model assumptions itself, e.g., Roche's description of the tidal deformation, limb-darkening and gravity-darkening laws. 
Also, the mutual irradiation formalism of \cite{Horvat-irradiation-effect} may not be adequate for a secondary component whose surface is very different from a spherical case. To express how highly inhomogeneous the surface of the secondary component can be, let us emphasize that our modelling suggests a difference in the temperature of the hottest and coolest places of around $500$\,K, i.e., $\sim 10$\% of $T_{\rm eff,2}$ (see Fig.~\ref{fig:phoebe-temperatures}). Moreover, our model neglects at least two phenomena that can have a non-negligible impact on the shape of the AB\,Cas's light curve, i.e. the Doppler beaming/boosting and the redistribution of heat in the atmosphere of the secondary component which is strongly illuminated by the much hotter primary ($T_{\rm eff,1}/T_{\rm eff,2}\approx 1.8$).
The cause of systematic differences between the model and the observations visible in the \textit{TESS} residuals may also be the presence of an accretion disk around the primary component, gas streams, and/or non-axisymmetric surface temperature distribution on the components connected with the occurrence of hot/cool spots. It seems reasonable to claim that while the mass transfer takes place in AB\,Cas, also a disk (or some scattered matter/stream) is present in this system. Due to the reflection and scattering of radiation, the disk can contribute to the phase-dependent modulation of the system's brightness. In turn, the accretion of this material onto the primary's surface may be local in certain favored astrographic latitudes and longitudes \citep[e.g.][]{Amman1973,Blondin1995,Piirola2005,Virnina2011}. This would give rise to hot spots on the surface that may be responsible for the additional flux variations seen in the residuals. Also, the presence of cool spots on the surface of secondary component caused by the magnetic field can not be ruled out either \citep[see e.g.][]{Cokluk2019}. Nevertheless, we are aware that all the phenomena described above can coexist and lead to systematic discrepancies between our model and observations.

\begin{table}
    \setlength{\tabcolsep}{3.0pt}
    \centering
    \footnotesize
    \caption{The comparison of results from the \texttt{PHOEBE\,2} and \citet{Hong2017} analysis.}
    \label{tab:PHOEBE_paramterers}
    \begin{tabular}{lll}

        \hline
        \hline
        & \texttt{PHOEBE\,2} Model   & \cite{Hong2017} \\
        & $(1\sigma)$   & $(1\sigma)$   \\
        \hline

        \multicolumn{3}{l}{\textbf{----- Orbital parameters -----}} \\
        Orbital period $P_{\rm orb}$\,(d)    & $1.3668810(6)^{\rm a,\star}$        & $1.3668918(2)$ \\
        Orbital inclination $i$\,($^\circ$) & $89.1(6)$ & $89.9(2)$ \\
        Moment of the superior & \multirow{2}{*}{$2458791.777388(20)$} & \multirow{2}{*}{$2452501.3463(3)$} \\
        conjunction $T_0$\,(BJD) &&\\
        $E(B-V)$\,(mag) & $0.182^{\rm b,\star}$ & $0.182(12)$ \\
        $M_2/M_1$                   & $0.184(12)^{\rm b,\star}$ & $0.184(12)$ \\

        \multicolumn{3}{l}{\textbf{----- Primary star (acceptor) -----}} \\
        Mass $M_1$\,(\Msun)           & $2.01(2)^{\rm b,\star}$ & $2.01(2)$ \\
        Radius $R_1$\,(R$_{\odot}$)  & $1.861(23)^{\rm c}$ & $1.84(2)$   \\
        $\log T_{\rm eff}$/K & $3.921(9)$ & $3.907(9)$ \\
        $\log L/L_{\odot}$      & $1.177(35)^{\dagger}$   & $1.12(4)$   \\
        $A_{\rm bol}$ & $1.0^\star$ & $1.0$ \\
        Gravity-darkening coefficient $g$ & $1.0^\star$ & $1.0$ \\

        \multicolumn{3}{l}{\textbf{----- Secondary star (donor) -----}} \\
        Mass $M_2$\,(\Msun)           & $0.37(2)^{\rm b,\star}$ & $0.37(2)$ \\
        Radius $R_2$\,(R$_{\odot}$)  & $1.692(31)^{\rm c,d}$ & $1.69(3)$   \\
        $\log T_{\rm eff}$/K & $3.677(6)$ & $3.692(13)$ \\
        $\log L/L_{\odot}$      & $0.117(27)^{\dagger}$   & $0.17(5)$ \\
        $A_{\rm bol}$ & $0.66(8)$ & $0.5$ \\
        Gravity-darkening coefficient $g$ & $0.44(29)$ & $0.32$ \\

        \hline

        \multicolumn{3}{l}{\makecell[l]{\textbf{Notes: }\\$^{\star}$ Fixed during the fitting procedure, \\
        $^{\rm a}$ Obtained from the Fourier analysis of the \textit{TESS} light curve, \\
        $^{\rm b}$ After \cite{Hong2017}, \\
        $^{\rm c}$ The value refers to the so-called equivalent radius, $R_{\rm equiv}$. Note that it is \\ generally different from the radius defined in \texttt{MESA}, \\
        $^{\rm d}$ Assuming that during the modelling the semi-detached geometry of the \\system is always preserved, i.e.~the secondary component fills its Roche lobe, \\
        $^{\dagger}$ From $\log L/L_{\odot} = 4 \log(T_{\rm eff}/T_{\rm eff \odot}) + 2 \log(R/R_{\odot})$.}}
    \end{tabular}
\end{table}

\section{Change of the orbital period}
\label{sec:orb_period_variation}

Large-scale effects, especially the mass transfer and changes in the orbital momentum, that occur in the system during its semi-contact phase of evolution may have a significant impact on the system's geometry and its orbital characteristics. One of the evidence of the ongoing MT is the change in the orbital period.
In order to examine this phenomenon and determine if it is present in the AB\,Cas system, we used all the historical times of light minima gathered in the O-C Gateway\footnote{http://var2.astro.cz/ocgate/} and we analysed them by means of the $O-C$ diagrams. These data contain 760 times of primary light minima spread over 92 years, between June 1928 and August 2020.
In the first step, the binary light curve model calculated in Sec.\,\ref{sec:binarymodelling} was used as a template, with the reference time of first \textit{TESS} minimum light, $T_{\rm ref} \equiv T_0 = \rm BJD\ 2458791.777388$\,d. As a referential value of the orbital period we adopted the solution from the light curve modelling, i.e. $P_{\rm ref} \equiv P_{\rm orb} = 1.3668810$\,d.
Next, using the ephemeris $T=T_{\rm ref} - E \times P_{\rm ref}$ the reference time was transferred to the time of minimum light closest to the measured times gathered from the literature by subtracting  or adding the integer number of orbital cycles, $E$.
The difference between the observed time of minimum light and the ephemeris described above returns the values of $(O-C)$.

\begin{figure}
    \centering
    \includegraphics[width=\columnwidth,clip]{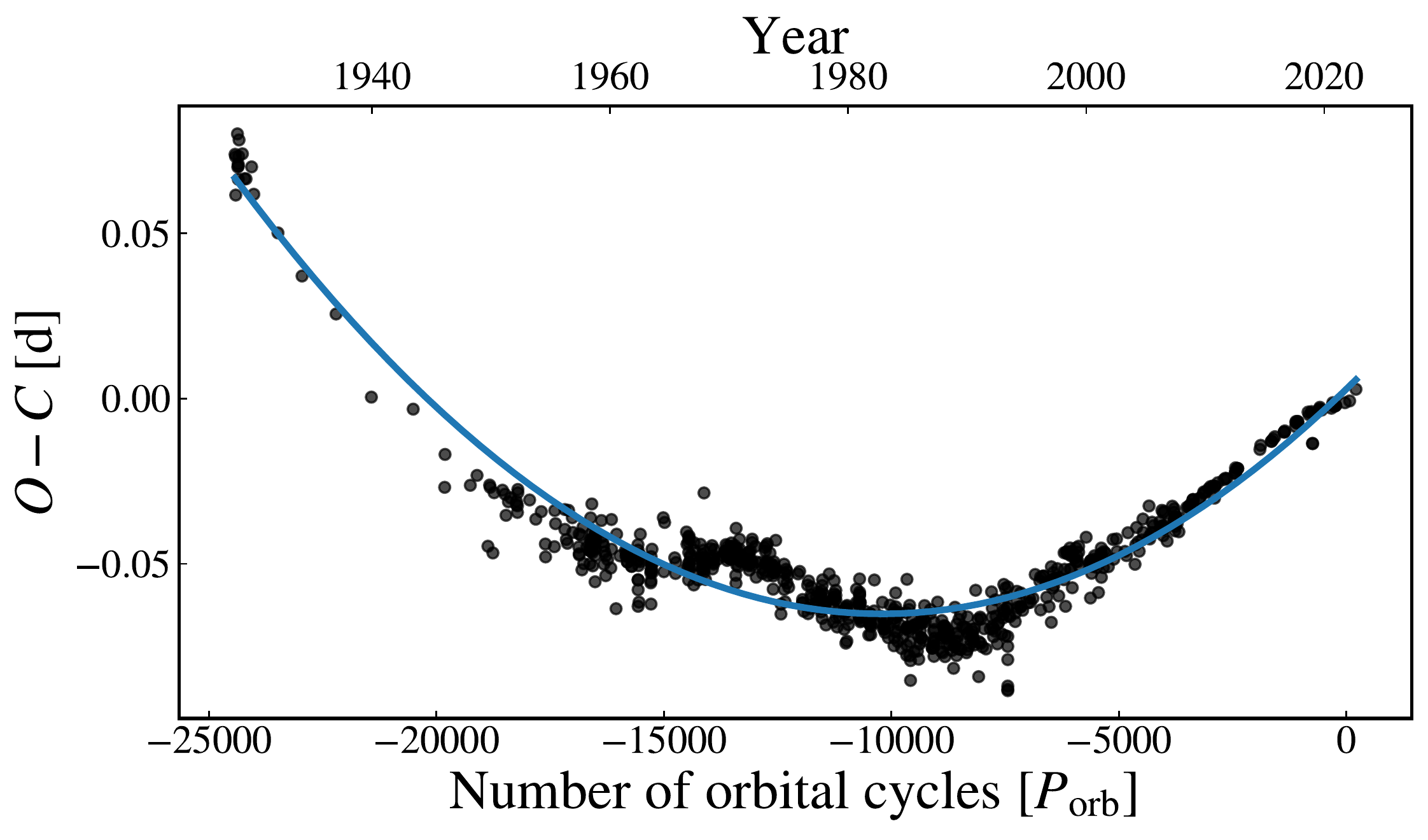}
    \caption{The $O-C$ diagram for the AB\,Cas system. The fitted parabolic curve is shown with a blue line.}
    \label{img:P_orb_change}
\end{figure}

The results of the period change analysis are presented in Fig.\,\ref{img:P_orb_change}.
The overall changes in $(O-C)$ resemble the parabolic variation, therefore
we fitted the second order polynomial, in the form of $(O-C)(E) = aE^2 + bE + c$, to them by means of the linear least-squares method fit and found the following formula:
\begin{align}
(O-C)(E)      & = 6.524(8) \times 10^{-10}\ E^2 \nonumber \\
            & + 1.332(2) \times 10^{-5}\ E    \nonumber \\
            & + 2.866(996) \times 10^{-3},  \label{eqn:Tmin}
\end{align}
where $E$ is the number of orbital cylces that have elapsed from $T_{\rm ref}$.
As a result, the coefficient in the quadratic term was found to be $a=6.524 \times 10^{-10}$\,d 
and the rate of orbital period change was calculated as
\begin{equation}
\dot{P}_{\rm orb} \equiv \frac{\mathrm{d} P_{\rm orb}}{\mathrm{d}t} = \frac{2a}{P_{\rm ref}} = 0.0301(4) \rm\ s/yr.
\label{eqn:dotP}
\end{equation}
This result qualitatively agrees with the result obtained by \cite{Soydugan2003}, i.e., $\dot{P}_{\rm orb}=0.03\ \rm s/yr$, however this value is slightly lower than the value obtained by \cite{Abedi2007}, whom value was determined on $\dot{P}_{\rm orb}=0.05\ \rm s/yr$.

Apart from the parabolic variation, data in Fig.\,\ref{img:P_orb_change} show additional bump at $E\sim -13000$. This bump is likely to be connected with the sudden change in the MT rate or with the light travel-time effect caused by a third body orbiting around the center of mass. Indeed, the signatures of a third body have been reported in the past by \cite{Soydugan2003} and \cite{Abedi2007}, who have investigated this phenomenon more closely obtaining a star with a minimum mass of $0.2$\,\Msun\ orbiting the system's barycenter with the orbital period longer than 25 years.

\section{Frequency analysis}
\label{sec:freqanalysis}

To extract the pulsational characteristics of the system we followed the Fourier analysis for the light curve corrected for the binary orbit. For this purpose, we used the residuals from all three \textit{TESS} sectors that were obtained by subtracting the \PHOEBE\ model from the data.

We calculated the amplitude spectra employing a discrete Fourier transform
\citep{Deeming1975,Kurtz1985} and followed the standard pre-whitening procedure.
The periodograms were calculated up to the pseudo-Nyquist frequency, i.e. $f_{\rm N} \sim 360$\,\cpd.
We assumed the signal-to-noise ratio limit $\rm S/N=4$ as a threshold for significant frequencies \cite[see][]{Breger1993,Kuschnig1997}.
The noise was calculated as an average amplitude value in a 1\,\cpd\ window centred on a given peak before its extraction.

Despite of subtracting the orbital model from the data, we noticed the presence of the integer multiples of the orbital frequency in the periodograms.
These signals can occur due to trends in the residuals (see Fig.\,\ref{fig:phoebe-fit}b) and due to some difficulties during the light curve normalisation procedure, since the depths of eclipses are slightly different in each sector.
To correct the residuals for these signals, we additionally corrected separately each sector for 100 orbital harmonics.

\begin{figure}
    \centering
    \includegraphics[width=\columnwidth,clip]{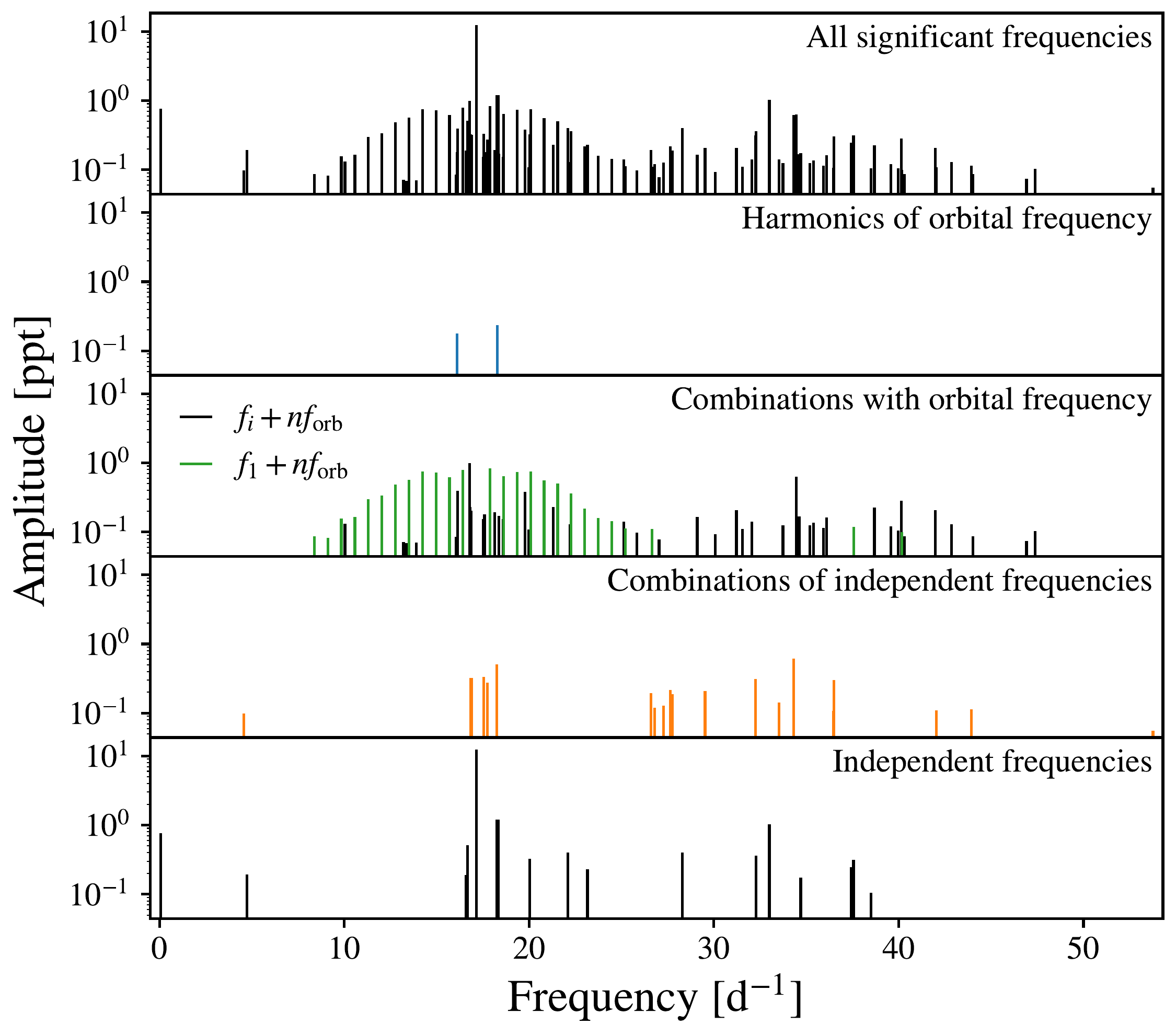}
    
    \caption{The Fourier spectrum for the AB\,Cas system.
    In the top panel, we show all frequencies that were found in the \textit{TESS} data.
    Blue lines on the second from the top panel mark orbital frequency harmonics, whereas the combinations with the orbital frequency are presented in the middle panel. 
    Green lines mark combinations of $f_1$ with the orbital frequency, whereas the black lines mark combinations of other pulsational frequencies with the orbital frequency.
    The penultimate panel contains combinations, and the last panel presents only independent frequency peaks.
    }
    \label{img:osc_freq}
\end{figure}

Careful analysis revealed 114 frequency peaks from the \textit{TESS} data with the most prominent peak at $f=17.156430(3)$\,\cpd.
We adopted the 1.5 Rayleigh limit ($1.5/T$) as a resolution criterion \citep{Loumos1978}, obtaining $1.5\Delta f_{\rm R} \equiv 1.5/T = 0.00686$\,\cpd. We checked whether some of the found frequencies are separated by the distance lower than the $1.5\Delta f_{\rm R}$. Whenever we found such a pair of frequencies, we checked their amplitudes and removed the one with the lower amplitude. During this procedure, we rejected two frequencies. The remaining set of 112 significant frequencies we regard as a final one for the further identification of possible combinations. We show these frequencies in the top panel of Fig.\,\ref{img:osc_freq}.
We note that the dominant peak from the analysed \textit{TESS} periodograms agrees with the main frequency found by \cite{Rodriguez2004}.
Moreover, we found frequencies $f_3$ and $f_{22}$ which are close to the secondary peak reported by \cite{Rodriguez2004}, $f=18.25$\,\cpd. Taking into account that the amplitude of $f_3$ is an order of magnitude higher than for $f_{22}$, we accept $f_3$ as an equivalent of $f_2$ from \cite{Rodriguez2004}.
The discrepancy ($\sim 0.008$\,\cpd) may results from the fact that this frequency is barely detectable in the \textit{uvby} data.
The tertiary peak from \cite{Rodriguez2004}, $f=34.64$\,\cpd, matches $f_{98}$ resulting from our analysis.

Using a simple method of finding combination frequencies ($m \times f_i + n \times f_j,\ m,n \in [-10,10]$), orbital harmonics ($N \times f_{\rm orb}$, where $N \in \mathbb{N}^+$ and $ N \le 100$) and combinations with the orbital frequency ($m \times f_i + n \times f_{\rm orb},\ m \in [-10,10]$ and $n \in [-20,20]$) we determined, that 17 amongst all of the observed frequencies seem to be independent with the accuracy of the adopted Rayleigh resolution, $1.5\Delta f_{\rm R}$. We show all possible classifications of the observed frequencies in Fig.\,\ref{img:osc_freq}.

Even though after subtracting the \PHOEBE\ model from the data we additionally corrected each sector for 100 orbital harmonics, there are still visible two signals matching the orbital harmonics. The presence of these harmonics may indicate that the depths of eclipses vary even within one sector. It can be explained at least in two ways.
Firstly, the depths of eclipses can be a manifestation of some observational uncertainties introduced by \textit{TESS} detector.
Secondly, in order to improve the light curve modelling (see Sect.\,\ref{sec:binarymodelling_preparations}) from the \textit{TESS} and the \textit{uvby} data sets we subtracted the dominant frequency. This could introduce the additional signals in the primary eclipses, where the pulsations are less visible than in the moments of quadratures. These artefacts, in turn, can be responsible for orbital harmonics visible in the data.

We also found frequencies that seem to be resulting from possible combinations with the orbital frequency. Such behaviour is known to occur in the binary frequency spectra, as the orbital movement of the pulsator causes systematic shifts of frequencies due to the Doppler effect \citep{Shibahashi2012}. 
However, in the case of AB\,Cas this effect is negligible.
It is also expected to find such frequencies when analysing data from eclipsing binaries, since the component’s contribution to the total light changes with the orbital phase, especially during eclipses.
Moreover, the combinations with the orbital frequency could appear due to some geometric effects, implying that the star pulsates as a tilted-pulsator. However, we found no amplitude and phase modulation of the pulsating frequencies with the orbital phase, which excludes such origin of these signals.
Finally, such harmonics can result from the effects of the tidal force in a close binary that introduces another axis of symmetry. This leads to the splitting of the mode frequency into equidistant frequencies spaced by multiples of the orbital frequency. To first order, the radial mode frequency remains unaffected by the tidal force \citep{Reyniers2003a,Reyniers2003b,Balona2018,Steindl2021}. In our case, the structure of equidistant frequencies is the most prominent for the case of $f_1$. As $f_1$ is the radial mode, hence no splitting is possible.

The mode identification for pulsations occurring in the AB\,Cas primary is available in the literature solely for the dominant signal, $f_1$. This is an obvious consequence of the relatively high amplitude of the peak, compared to the lower-amplitude signals found by \cite{Rodriguez2004}, i.e., $f_2 \approx 18.25$\,\cpd\ and $f_3 \approx 34.64$\,\cpd. This is also the only frequency with the amplitude and phase values determined in the literature.
The analyses claiming the primary frequency is a radial mode were made independently by \cite{Rodriguez1998,Rodriguez2004} and by \cite{Daszynska2003}. These analyses were based on different approaches, i.e. spatial filtering and photometric amplitudes and phases, however, the results unequivocally point to  $\ell=0$. 
Moreover, the calculations of the pulsational constant Q \citep{Rodriguez1998}, following the \cite{Fitch1981} and \cite{Breger1990} method, indicates that the $f_1$ is a radial fundamental mode.

Given, that the dominant frequency is most likely the fundamental radial mode, we checked the frequency ratios to identify possible higher radial overtones. We used the theoretical frequencies of the radial pulsations calculated for the purpose of Sect.\,\ref{sec:SeismicModelling}.
The frequency ratio for the fundamental and first overtone radial mode was adopted in
the range $f_{\rm F}/f_{\rm 1O} \in [0.7732, 0.7747]$. We also checked the possibility of the presence of higher overtones, with the allowed ratio ranges of: $f_{\rm F}/f_{\rm 2O} \in [0.6327, 0.6337]$, $f_{\rm F}/f_{\rm 3O} \in [0.5325, 0.5332]$ and $f_{\rm F}/f_{\rm 4O} \in [0.4558, 0.4563]$. 
Taking into account the above mentioned ratios, in the whole observed range of frequencies we found no possible radial overtones.

In Appendix\,\ref{sec:appendix} we provide a complete list of the significant frequencies found in the data. We maintained the original numeration from the pre-whithening procedure. Possible combinations are listed in the Remarks column.

\section{Binary-evolution models}
\label{sec:binaryevolution}

From a vast number of known eclipsing binary systems \citep[e.g.,][]{Prsa2011,LaCourse2015,Kirk2016,Prsa2021}
there are only a few well-studied close binaries that contain \dsct\ components, e.g., TT\,Hor \citep{Streamer2018} and KIC\,10661783 \citep{Miszuda2021}.
These systems, thanks to the detailed evolutionary modelling, have precise estimates of the initial parameters that allow to reconstruct their evolution.
Inclusion of the binary interactions into the modelling is, thus, crucial to understand the fundamental processes that took place in the past and that led the system to evolve into its current state.
Also, as was shown by \cite{Miszuda2021} for KIC 10661783, the
binary evolution can help to explain the excitation of high-order g modes in \dsct\ stars.

To model the AB\,Cas binary not as two isolated stars, but as an effect of the binary interactions in the past,
we used the \texttt{MESA} code \citep[Modules for Experiments in Stellar Astrophysics,][version r12115]{Paxton2011, Paxton2013, Paxton2015, Paxton2018, Paxton2019}, with the \texttt{MESA-binary} module. \texttt{MESA} relies on the variety of the input microphysics data.
The \texttt{MESA} EOS is a blend of the OPAL \citep{Rogers2002}, SCVH \citep{Saumon1995}, FreeEOS \citep{Irwin2004}, HELM \citep{Timmes2000}, and PC \citep{Potekhin2010} EOSes.
Radiative opacities are primarily from the OPAL project \citep{Iglesias1993,Iglesias1996}, with data for lower temperatures from \citet{Ferguson2005} and data for  high temperatures, dominated by Compton-scattering from \citet{Buchler1976}. Electron conduction opacities are from \citet{Cassisi2007}.
Nuclear reaction rates are from JINA REACLIB \citep{Cyburt2010} plus additional tabulated weak reaction rates from \citet{Fuller1985}, \cite{Oda1994} and \cite{Langanke2000}. Screening is included via the prescription of \citet{Chugunov2007}. Thermal neutrino loss rates are from \citet{Itoh1996}.
The \texttt{MESA-binary} module allows to construct a binary model and to evolve its components simultaneously, considering several important interactions between them.
In particular, this module incorporates angular momentum evolution due to the mass transfer.
Roche lobe radii in binary systems are computed using the fit of \citet{Eggleton1983}. Mass-transfer rates in Roche lobe overflowing binary systems are determined following the prescriptions of \citet{Ritter1988} and \citet{Kolb1990}.

In our evolutionary computations, we used the AGSS09 \citep{Asplund2009} 
initial chemical composition of the stellar matter and the OPAL opacity tables. We adopted the Ledoux criterion for the convective instability with the mixing-length theory description by \cite{Henyey1965} and the semi-convective mixing with $\alpha_{\rm SC}=0.01$. The diffusive exponential overshooting scheme was applied with the free parameter $f_{\rm ov}$ \citep{Herwig2000}.
For the large-scale effects we used the mass transfer of Kolb's type \citep{Kolb1990} and included the stellar winds from both components following the prescription of \cite{Vink2001}. For the sake of simplicity of computations we assumed a constant eccentricity throughout the system's evolution, i.e., $e=0$, ignored the rotation of stars and disabled the tides.

In order to reproduce the evolution of the system, we built an extensive grid of evolutionary models. To do that, we constructed a set of varying parameters like the initial orbital period, initial masses of the components, the metallicity and the initial hydrogen abundance, overshooting from the convective core and a fraction of the mass lost during the mass transfer with the ranges as given in Table\,\ref{tab:MESA_parameter_ranges}. From this set we drawn 50\,000 vectors of initial parameters and calculated the evolutionary tracks assuming a value of the mixing-length theory parameter, $\alpha_{\rm MLT}$. This parameter defines the efficiency of convection.
We tested the values of $\alpha_{\rm MLT}$ from $0.0$ up to $1.8$, with the step of $\Delta \alpha_{\rm MLT}=0.1$. For each $\alpha_{\rm MLT}$, we have calculated 50\,000 models, resulting in a total of $\sim 950\,000$ models.

For each grid we let the parameters that characterise the binary evolutionary tracks to be randomly chosen from a uniform distribution within the given ranges. All of these parameters were chosen independently of others, except for the masses.
We adopted, that the total initial mass of the system cannot be less than the observed value, that is 2.38\,\Msun\ \citep{Hong2017}.
However, the mass transfer does not necessarily have to be conservative, so we set 3.00\,\Msun\ as an upper mass limit.
As a limit on a single component's mass we set the mass range between 0.5 to 2.5\,\Msun. Each time the masses were controlled to ensure that the total initial mass of the system was in the range from 2.38\,\Msun\ to 3.00\,\Msun\ and that the initial mass of the donor is larger than acceptor. The initial orbital period was considered in the range from 1.2 to 5 days. The allowed steps were as small as $\Delta M_{\rm don/acc, ini} = 10^{-3}$\,\Msun\ and $\Delta P_{\rm ini} = 10^{-5}$ days.
We also tested the mass-transfer conservativity rate from $\beta = 0.0$ up to $\beta = 0.3$ with the step $\Delta \beta = 10^{-2}$. The $\beta$ parameter denotes a fraction of mass that is lost from the vicinity of acceptor in the form of a fast wind during the MT \citep{Tauris2006}. The other parameters describing the mass-transfer conservativity as $\alpha$ (fraction of mass lost from the vicinity of the donor as fast wind), $\delta$ (fraction of mass lost from circumbinary coplanar toroid) and $\gamma$ (radius of the circumbinary coplanar toroid) were all set to default values, i.e. $\alpha,\beta,\gamma = 0$.
The models have different values of metallicity $Z$, initial hydrogen abundance $X_0$ and overshooting $f_{\rm ov}$ in the ranges: $Z \in [0.013,0.026]$, $X_0 \in [0.680,0.740]$ and $f_{\rm ov} \in [0.000,0.030]$ with the values chosen with the accuracy of $\Delta = 10^{-3}$.
A summary of considered parameter ranges during the evolutionary modelling is presented in Table\,\ref{tab:MESA_parameter_ranges}. 

\begin{table}
    \setlength{\tabcolsep}{3.5pt}
    \centering
    \caption{A summary of considered parameter ranges during the evolutionary modelling of the AB\,Cas system. The following columns contain: a description of the parameters, their minimum and maximum values, and the accuracy with which the parameters were drawn from the uniform distributions.}
    \label{tab:MESA_parameter_ranges}
    \begin{tabular}{lccc}

        \hline
        \hline
        Parameter & Min & Max & Step $\Delta$ \T\B \\
        \hline

        Initial orbital period, $P_{\rm ini}\,\rm [d]$               & 1.2   & 5.0   & $10^{-5}$\T\\
        Initial donor mass, $M_{\rm don,ini}\,\rm[M_{\odot}]$       & 0.5   & 2.5   & $10^{-3}$  \\
        Initial acceptor mass, $M_{\rm acc,ini}\,\rm[M_{\odot}]$    & 0.5   & 2.5   & $10^{-3}$  \\
        Metallicity, $Z$                                            & 0.013 & 0.026 & $10^{-3}$  \\
        Initial hydrogen abundance, $X_0$                           & 0.68  & 0.74  & $10^{-3}$  \\
        Mixing-length theory parameter, $\alpha_{\rm MLT}$          & 0.0   & 1.8   & $10^{-1}$  \\
        Convective-core overshooting, $f_{\rm ov}$                  & 0.00  & 0.03  & $10^{-3}$  \\
        Fraction of mass lost during MT, $\beta$                    & 0.0   & 0.3   & $\ 10^{-1}$ \B\\
        \hline

    \end{tabular}
\end{table}

From all calculated evolutionary tracks we selected the models that exactly reproduce the observed value of $P_{\rm orb}$, i.e. $P_{\rm orb}~\equiv~P_{\rm orb}^{\rm obs}~=~1.3668810$\,d. Next, for each model we calculated a discriminant $D^2$ defined as:
\begin{align}
D^2 & \equiv \frac{1}{2}\sum_{\rm i=1}^{\rm 2} \left( \frac{X_{\rm obs, i}-X_{\rm model, i}}{3\sigma_{\rm obs, i}} \right)^2.  \label{eqn:D2}
\end{align}
Here, $X$ denotes the mass or radius of a given star and  $\sigma$ stands for the respective error of these parameters. 
Then, the mean value of $D^2$ was computed as
\begin{align}
< D^2 > = \frac{D_1^2 + D_2^2}{2},  \label{eqn:D2mean}
\end{align}
where $D_1^2$ and $D_2^2$ are the discriminants for the primary and secondary, respectively.

\begin{figure*}
    \centering
    \flushright
    \includegraphics[width=\textwidth,clip]{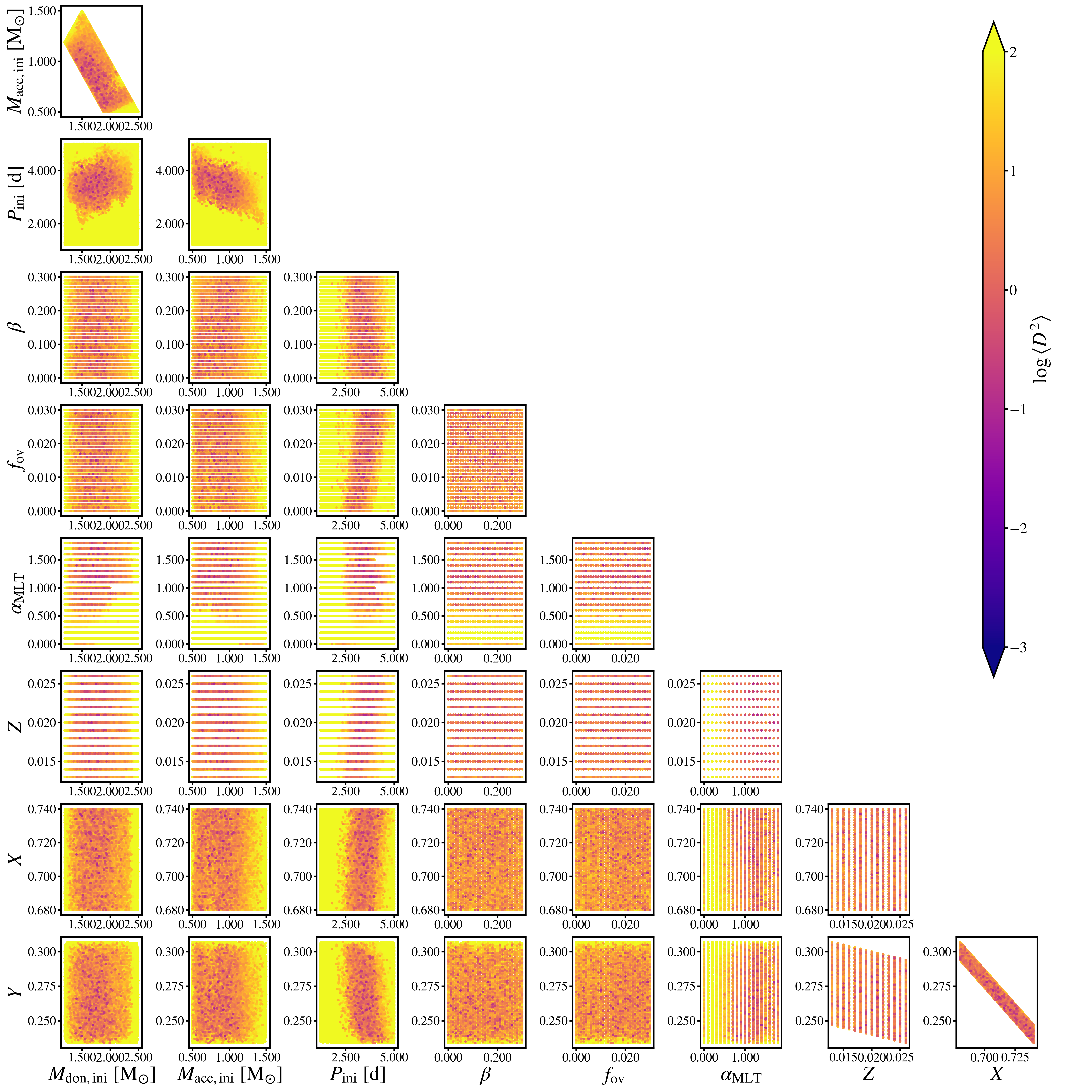}
    \caption{Corner plots showing the dependencies between the considered parameters.
    The value of $<D^2>$ is coded with colors.
    }
    \label{MESA:corner_plots}
\end{figure*}

The values of $<D^2>$ coded with colors are presented on the corner plots, in Fig.\,\ref{MESA:corner_plots}.
As one can see, some correlations between the parameters exist, e.g. $P_{\rm ini}$ versus $M_{\rm acc,ini}$ or $M_{\rm don,ini}$, and between $f_{\rm ov}$, $Z$ and $X_0$ with the value of $P_{\rm ini}$. Our results prefer the value of $\alpha_{\rm MLT}$ greater than 0.7 and the value of $P_{\rm ini}$ greater than 2.5\,d. 
None of the explored values of $\beta$, however, are preferred.

From the computed grid, we selected models, that for the observed value of $P_{\rm ini}$ reproduce masses and radii of both components within their $3\sigma$ errors.
From the total of $950\,000$ evolutionary tracks, we found 58 models that met our selection criteria. These models are summarised in Table\,\ref{tab:fitting_models}, which includes the initial parameters used to compute the evolutionary tracks and the final parameters of the models, like masses, radii, effective temperatures and luminosities.
However, not all of these models fit in the allowed range of the effective temperatures determined from the \PHOEBE\ analysis (see Table\,\ref{tab:PHOEBE_paramterers}). 
There are thirteen models that reproduce also the values of $T_{\rm eff}$ within the $3\sigma$ error. These are the models with numbers: 6, 12, 13, 16, 21, 37, 38, 39, 40, 42, 52, 53 and 57.
These models indicate the age of AB\,Cas to be between $1.86 - 4.26$\,Gyr and remaining parameters in the following ranges:
$P_{\rm ini} \in[3.04530, 4.07314]$\,d, $M_{\rm don,ini} \in [1.573, 2.042]$\,\Msun, $M_{\rm acc,ini} \in [0.589, 1.033]$\,\Msun, $f_{\rm ov} \in [0.009, 0.028]$, $\alpha_{\rm MLT} \in [1.1, 1.6]$ and $\beta \in [0.02, 0.28]$. Using our approach we were not able to constrain the chemical composition. Our models cover the whole adopted range for the initial hydrogen abundance $X_0 \in [0.68, 0.74]$ and the metallicity $Z \in [0.015, 0.025]$.

\begin{figure*}
    \begin{tabular}{cc}
        \includegraphics[width=1.0\columnwidth,clip]{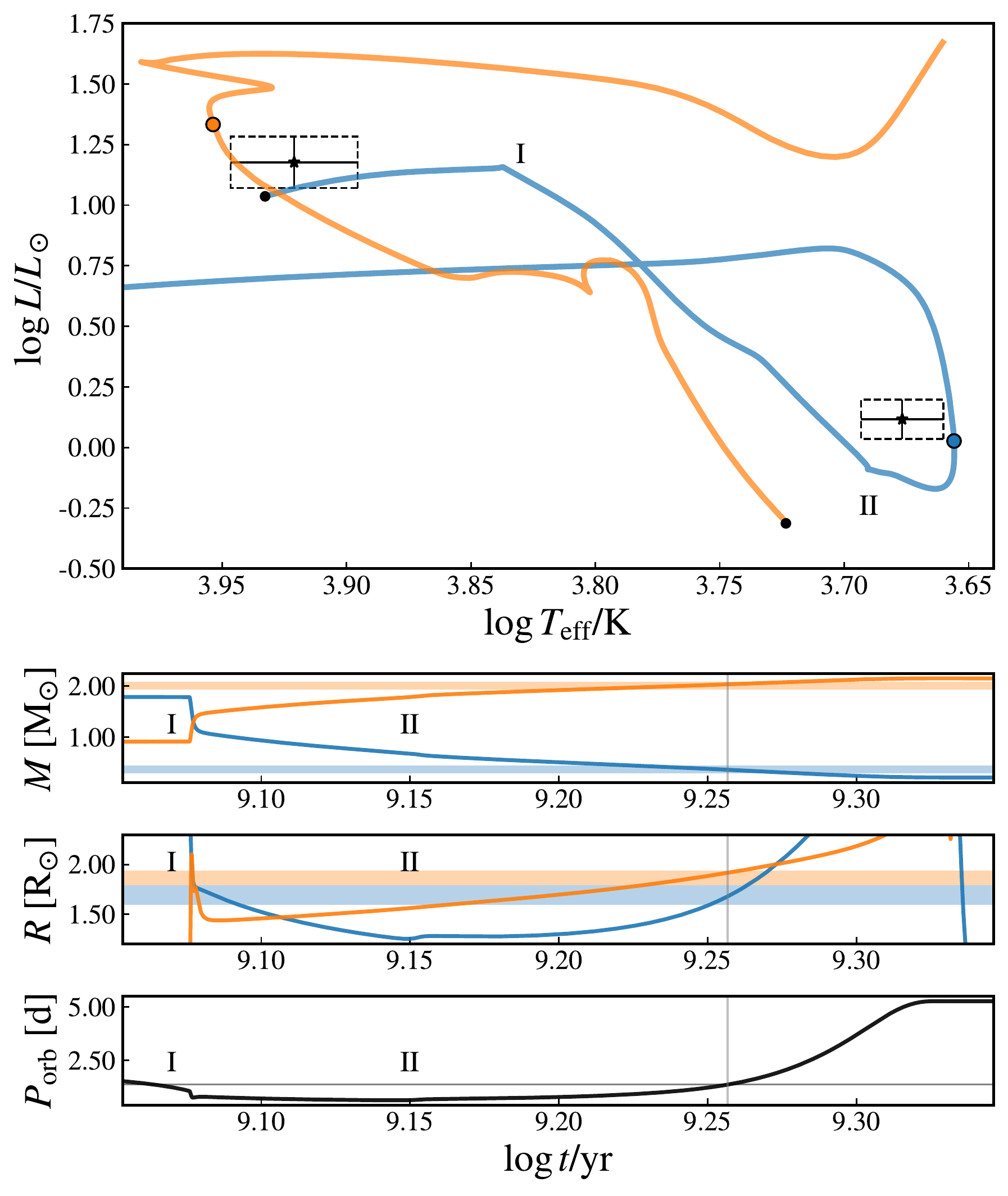} &
        \includegraphics[width=1.0\columnwidth,clip]{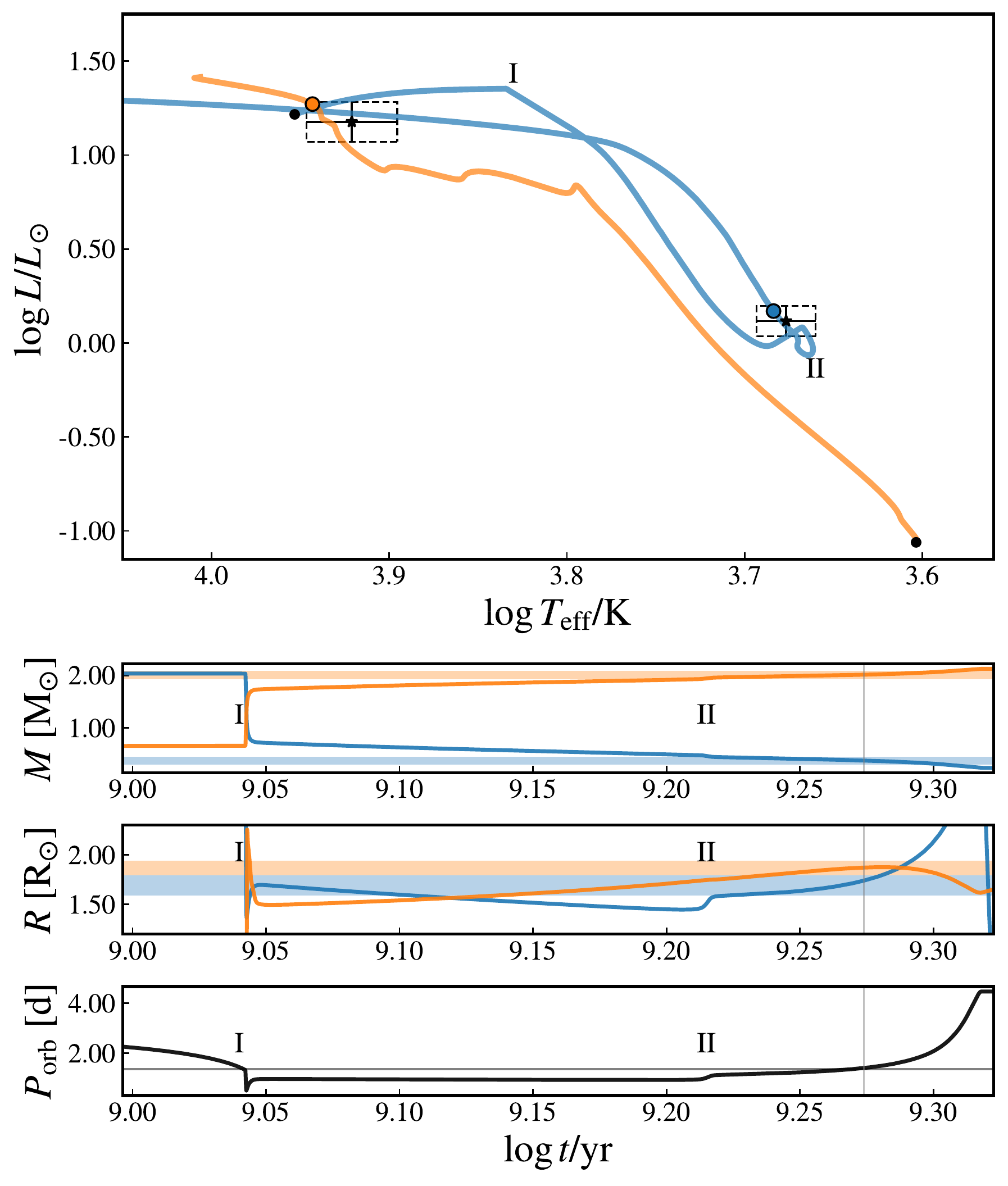}
    \end{tabular}
    \caption{The HR diagrams and the evolution of masses, radii and orbital period for models No.\,1 and No.\,13 (see Table\,\ref{tab:fitting_models}) on the left and right panels, respectively. Orange and blue lines show the evolutionary tracks for the acceptor and donor, respectively. The observed positions are marked with the error boxes. \texttt{I} shows the onset of the mass transfer and \texttt{II} marks the core conversion to fully helium and the beginning of the shell H-burning for the donor star.}
    \label{MESA:best_models}
\end{figure*}

From all the evolutionary models, we chose models No.\,1 and No.\,13 to show the evolutionary tracks, along with the evolution of masses, radii and orbital period in Fig.\,\ref{MESA:best_models}.
The top panel shows the HR diagrams with blue and orange lines for the evolutionary tracks of donor and acceptor, respectively. The black dots mark the Zero-Age Main Sequence (ZAMS) and the boxes show the observed position of the components with $3\sigma$ errors. Lower panels, from the top to bottom, show the evolution of masses, radii and orbital period. The horizontal stripes mark the observed ranges of parameters with the $3\sigma$ errors and the vertical grey lines indicate the model that for the observed value of $P_{\rm orb}$ reproduces masses and radii of the components. The position of these models is marked on each evolutionary track in the HR diagram with orange and blue dots for the acceptor and donor, respectively. Moreover, we mark two moments in the evolution of the considered models: \texttt{I} presents the onset of the mass transfer and \texttt{II} marks the core conversion to fully helium and the beginning of the shell H-burning for the donor star. This moment corresponds roughly to the minimum of the effective temperature and luminosity of the donor.

For each of the selected evolutionary models obtained during our computations, the formation scenarios resemble the formation of EL\,CVn-type binaries \citep[e.g.,][]{Maxted2014,Chen2017,Miszuda2021} in which the donor looses its outer layers in a relatively high rate, preventing it to reach the RGB phase. Instead, it is stripped of its outer layers leaving only a fraction of its initial hydrogen shell \citep[0.001-0.005\Msun,][]{Maxted2014B}. That leads the donor to the nearly-constant luminosity phase and later towards the thin-layer instability. 
In this stage mass of the hydrogen layer is being substantially reduced by the unstable H-burning via CNO cycle that leads the star to the cooling sequence of the helium white dwarfs.

\begin{table*}
    \setlength{\tabcolsep}{3.5pt}
    \centering
    \scriptsize
    \caption{The results of evolutionary computations summarising the initial and final parameters of models reproducing the system at its current state.}
    \label{tab:fitting_models}
    \begin{tabular}{rcccccccc|ccccccccc}

    \hline
    \hline
    &\multicolumn{8}{c}{|--------------------------------- Initial parameters --------------------------------------|} & \multicolumn{9}{c}{|----------------------------------------------------- Final parameters ---------------------------------------------------|} \\
    $N_{\rm mod}$ &$P_{\rm ini}$ & $M_{\rm don,ini}$ & $M_{\rm acc,ini}$ & $Z$ & $X_0$ & $\alpha_{\rm MLT}$ & $f_{\rm ov}$ & $\beta$ & $M_{\rm acc}$ & $R_{\rm acc}$ & $\log T_{\rm eff}^{\rm acc}$ & $\log L_{\rm acc}/L_{\odot}^{\dagger}$ & $M_{\rm don}$ & $R_{\rm don}$ & $\log T_{\rm eff}^{\rm don}$ & $\log L_{\rm don}/L_{\odot}^{\dagger}$ & $\rm Age$ \T\\
    &$\rm [d]$ & $\rm[M_{\odot}]$ & $\rm[M_{\odot}]$ &  &  &  &  &  & $\rm[M_{\odot}]$ & $\rm[R_{\odot}]$ &  &  & $\rm[M_{\odot}]$ & $\rm[R_{\odot}]$ & & & $\rm [Gyr]$  \B\B\\
    \hline
$1$ & $2.62267$ & $1.791$ & $0.911$ & $0.015$ & $0.711$ & $1.6$ & $0.003$ & $0.21$ & $2.038$ & $1.917$ & $3.954$ & $1.333$ & $0.360$ & $1.682$ & $3.656$ & $\ \ \,0.029$ & $1.81$\T\\
$2$ & $2.92500$ & $1.912$ & $0.855$ & $0.022$ & $0.709$ & $1.0$ & $0.001$ & $0.23$ & $2.013$ & $1.927$ & $3.919$ & $1.198$ & $0.403$ & $1.747$ & $3.602$ & $-0.156$ & $1.82$ \\
$3$ & $2.94498$ & $1.726$ & $0.836$ & $0.018$ & $0.709$ & $1.2$ & $0.003$ & $0.08$ & $2.052$ & $1.834$ & $3.952$ & $1.288$ & $0.400$ & $1.742$ & $3.622$ & $-0.078$ & $2.05$ \\
$4$ & $3.00855$ & $1.650$ & $0.980$ & $0.014$ & $0.725$ & $1.3$ & $0.014$ & $0.19$ & $1.980$ & $1.861$ & $3.951$ & $1.295$ & $0.409$ & $1.754$ & $3.657$ & $\ \ \,0.068$ & $2.64$ \\
$5$ & $3.01036$ & $1.648$ & $1.052$ & $0.019$ & $0.726$ & $1.1$ & $0.005$ & $0.26$ & $1.975$ & $1.883$ & $3.921$ & $1.186$ & $0.395$ & $1.735$ & $3.608$ & $-0.137$ & $2.74$ \\
$6$ & $3.04530$ & $1.785$ & $0.962$ & $0.015$ & $0.734$ & $1.5$ & $0.015$ & $0.28$ & $1.959$ & $1.854$ & $3.938$ & $1.242$ & $0.393$ & $1.731$ & $3.679$ & $\ \ \,0.147$ & $2.46$ \\
$7$ & $3.08189$ & $1.888$ & $0.860$ & $0.017$ & $0.704$ & $1.8$ & $0.020$ & $0.24$ & $1.989$ & $1.913$ & $3.948$ & $1.308$ & $0.395$ & $1.734$ & $3.723$ & $\ \ \,0.322$ & $2.09$ \\
$8$ & $3.13209$ & $1.950$ & $0.594$ & $0.016$ & $0.722$ & $1.3$ & $0.004$ & $0.07$ & $2.057$ & $1.843$ & $3.956$ & $1.310$ & $0.372$ & $1.701$ & $3.656$ & $\ \ \,0.038$ & $1.62$ \\
$9$ & $3.17850$ & $1.610$ & $0.869$ & $0.019$ & $0.714$ & $1.2$ & $0.007$ & $0.16$ & $1.951$ & $1.852$ & $3.927$ & $1.198$ & $0.316$ & $1.612$ & $3.626$ & $-0.129$ & $2.93$ \\
$10$ & $3.19466$ & $1.697$ & $1.015$ & $0.019$ & $0.734$ & $1.4$ & $0.011$ & $0.26$ & $1.984$ & $1.849$ & $3.923$ & $1.181$ & $0.382$ & $1.715$ & $3.642$ & $-0.011$ & $2.85$ \\
$11$ & $3.20573$ & $1.544$ & $0.994$ & $0.022$ & $0.693$ & $1.0$ & $0.009$ & $0.13$ & $1.973$ & $1.901$ & $3.926$ & $1.215$ & $0.413$ & $1.761$ & $3.604$ & $-0.139$ & $3.05$ \\
$12$ & $3.21384$ & $1.787$ & $0.819$ & $0.018$ & $0.711$ & $1.6$ & $0.014$ & $0.14$ & $2.028$ & $1.917$ & $3.944$ & $1.296$ & $0.375$ & $1.704$ & $3.681$ & $\ \ \,0.140$ & $2.32$ \\
$13$ & $3.21879$ & $2.042$ & $0.653$ & $0.016$ & $0.737$ & $1.3$ & $0.009$ & $0.18$ & $2.010$ & $1.866$ & $3.942$ & $1.263$ & $0.381$ & $1.713$ & $3.682$ & $\ \ \,0.150$ & $1.86$ \\
$14$ & $3.22468$ & $1.469$ & $1.114$ & $0.018$ & $0.737$ & $1.5$ & $0.017$ & $0.19$ & $1.989$ & $1.870$ & $3.929$ & $1.213$ & $0.382$ & $1.716$ & $3.642$ & $-0.008$ & $4.12$ \\
$15$ & $3.26320$ & $1.563$ & $0.992$ & $0.021$ & $0.719$ & $0.8$ & $0.007$ & $0.18$ & $1.953$ & $1.897$ & $3.912$ & $1.158$ & $0.385$ & $1.719$ & $3.579$ & $-0.260$ & $3.36$ \\
$16$ & $3.28919$ & $1.618$ & $0.970$ & $0.020$ & $0.699$ & $1.6$ & $0.020$ & $0.19$ & $1.967$ & $1.875$ & $3.934$ & $1.236$ & $0.380$ & $1.712$ & $3.674$ & $\ \ \,0.116$ & $2.96$ \\
$17$ & $3.30451$ & $1.642$ & $0.870$ & $0.014$ & $0.722$ & $1.3$ & $0.022$ & $0.09$ & $1.974$ & $1.809$ & $3.958$ & $1.299$ & $0.421$ & $1.772$ & $3.679$ & $\ \ \,0.166$ & $2.86$ \\
$18$ & $3.30713$ & $1.675$ & $1.025$ & $0.020$ & $0.701$ & $1.8$ & $0.027$ & $0.24$ & $1.985$ & $1.909$ & $3.934$ & $1.251$ & $0.405$ & $1.749$ & $3.697$ & $\ \ \,0.228$ & $2.88$ \\
$19$ & $3.31553$ & $1.518$ & $0.918$ & $0.025$ & $0.689$ & $1.3$ & $0.010$ & $0.10$ & $1.978$ & $1.875$ & $3.924$ & $1.193$ & $0.335$ & $1.642$ & $3.626$ & $-0.113$ & $3.41$ \\
$20$ & $3.31675$ & $1.621$ & $0.833$ & $0.015$ & $0.726$ & $1.6$ & $0.020$ & $0.03$ & $2.057$ & $1.863$ & $3.964$ & $1.351$ & $0.353$ & $1.671$ & $3.687$ & $\ \ \,0.148$ & $2.98$ \\
$21$ & $3.32623$ & $1.875$ & $0.821$ & $0.017$ & $0.740$ & $1.3$ & $0.014$ & $0.20$ & $1.972$ & $1.793$ & $3.932$ & $1.188$ & $0.430$ & $1.784$ & $3.662$ & $\ \ \,0.103$ & $2.31$ \\
$22$ & $3.35605$ & $1.764$ & $0.742$ & $0.015$ & $0.708$ & $1.7$ & $0.026$ & $0.08$ & $1.992$ & $1.794$ & $3.969$ & $1.337$ & $0.398$ & $1.738$ & $3.730$ & $\ \ \,0.354$ & $2.39$ \\
$23$ & $3.36406$ & $1.937$ & $0.624$ & $0.016$ & $0.685$ & $1.0$ & $0.029$ & $0.12$ & $2.004$ & $1.890$ & $4.002$ & $1.513$ & $0.355$ & $1.676$ & $3.783$ & $\ \ \,0.535$ & $2.28$ \\
$24$ & $3.38004$ & $1.889$ & $0.633$ & $0.019$ & $0.724$ & $1.0$ & $0.006$ & $0.05$ & $2.040$ & $1.841$ & $3.938$ & $1.234$ & $0.402$ & $1.746$ & $3.618$ & $-0.091$ & $1.95$ \\
$25$ & $3.38104$ & $1.558$ & $0.954$ & $0.024$ & $0.739$ & $1.2$ & $0.005$ & $0.12$ & $1.972$ & $1.806$ & $3.904$ & $1.082$ & $0.396$ & $1.735$ & $3.599$ & $-0.173$ & $3.77$ \\
$26$ & $3.38249$ & $1.933$ & $0.683$ & $0.015$ & $0.716$ & $1.0$ & $0.027$ & $0.13$ & $1.994$ & $1.892$ & $3.973$ & $1.399$ & $0.417$ & $1.768$ & $3.740$ & $\ \ \,0.410$ & $2.38$ \\
$27$ & $3.41461$ & $1.624$ & $0.984$ & $0.023$ & $0.726$ & $1.2$ & $0.011$ & $0.21$ & $1.969$ & $1.837$ & $3.912$ & $1.128$ & $0.371$ & $1.700$ & $3.619$ & $-0.108$ & $3.38$ \\
$28$ & $3.42054$ & $1.402$ & $1.022$ & $0.021$ & $0.713$ & $1.4$ & $0.021$ & $0.09$ & $1.980$ & $1.878$ & $3.929$ & $1.216$ & $0.342$ & $1.655$ & $3.644$ & $-0.031$ & $4.88$ \\
$29$ & $3.43473$ & $1.769$ & $0.732$ & $0.021$ & $0.707$ & $1.3$ & $0.010$ & $0.05$ & $2.035$ & $1.867$ & $3.938$ & $1.246$ & $0.392$ & $1.730$ & $3.643$ & $\ \ \,0.003$ & $2.33$ \\
$30$ & $3.48494$ & $1.854$ & $0.880$ & $0.019$ & $0.723$ & $1.7$ & $0.026$ & $0.24$ & $1.969$ & $1.793$ & $3.936$ & $1.205$ & $0.413$ & $1.760$ & $3.718$ & $\ \ \,0.316$ & $2.59$ \\
$31$ & $3.49906$ & $1.920$ & $0.596$ & $0.021$ & $0.688$ & $1.3$ & $0.013$ & $0.13$ & $1.959$ & $1.866$ & $3.940$ & $1.257$ & $0.346$ & $1.660$ & $3.700$ & $\ \ \,0.192$ & $2.08$ \\
$32$ & $3.52056$ & $2.040$ & $0.631$ & $0.025$ & $0.703$ & $1.3$ & $0.009$ & $0.23$ & $1.953$ & $1.844$ & $3.921$ & $1.167$ & $0.315$ & $1.610$ & $3.699$ & $\ \ \,0.161$ & $2.17$ \\
$33$ & $3.55639$ & $1.795$ & $0.847$ & $0.021$ & $0.702$ & $1.7$ & $0.029$ & $0.16$ & $2.039$ & $1.861$ & $3.959$ & $1.328$ & $0.366$ & $1.692$ & $3.746$ & $\ \ \,0.393$ & $2.82$ \\
$34$ & $3.55946$ & $1.651$ & $0.799$ & $0.021$ & $0.696$ & $1.2$ & $0.017$ & $0.04$ & $1.982$ & $1.841$ & $3.936$ & $1.229$ & $0.412$ & $1.760$ & $3.645$ & $\ \ \,0.026$ & $2.83$ \\
$35$ & $3.57138$ & $1.664$ & $0.918$ & $0.020$ & $0.721$ & $1.7$ & $0.028$ & $0.14$ & $2.036$ & $1.869$ & $3.943$ & $1.267$ & $0.356$ & $1.676$ & $3.707$ & $\ \ \,0.230$ & $3.42$ \\
$36$ & $3.61520$ & $1.935$ & $0.571$ & $0.018$ & $0.705$ & $1.2$ & $0.018$ & $0.04$ & $2.017$ & $1.832$ & $3.956$ & $1.302$ & $0.422$ & $1.774$ & $3.692$ & $\ \ \,0.219$ & $1.91$ \\
$37$ & $3.65843$ & $1.938$ & $0.761$ & $0.025$ & $0.713$ & $1.6$ & $0.017$ & $0.18$ & $2.010$ & $1.866$ & $3.919$ & $1.170$ & $0.407$ & $1.752$ & $3.685$ & $\ \ \,0.180$ & $2.30$ \\
$38$ & $3.66477$ & $1.573$ & $1.033$ & $0.023$ & $0.725$ & $1.5$ & $0.028$ & $0.17$ & $2.012$ & $1.905$ & $3.921$ & $1.197$ & $0.385$ & $1.720$ & $3.671$ & $\ \ \,0.107$ & $4.26$ \\
$39$ & $3.66552$ & $2.008$ & $0.635$ & $0.022$ & $0.705$ & $1.1$ & $0.018$ & $0.17$ & $1.961$ & $1.833$ & $3.932$ & $1.209$ & $0.402$ & $1.745$ & $3.691$ & $\ \ \,0.202$ & $2.16$ \\
$40$ & $3.69429$ & $2.013$ & $0.619$ & $0.022$ & $0.727$ & $1.2$ & $0.012$ & $0.15$ & $1.999$ & $1.852$ & $3.922$ & $1.178$ & $0.382$ & $1.716$ & $3.667$ & $\ \ \,0.091$ & $2.14$ \\
$41$ & $3.72727$ & $1.660$ & $0.879$ & $0.022$ & $0.738$ & $0.9$ & $0.015$ & $0.09$ & $2.034$ & $1.916$ & $3.917$ & $1.188$ & $0.384$ & $1.719$ & $3.605$ & $-0.157$ & $3.62$ \\
$42$ & $3.73838$ & $1.811$ & $0.593$ & $0.022$ & $0.682$ & $1.3$ & $0.018$ & $0.05$ & $1.952$ & $1.816$ & $3.942$ & $1.241$ & $0.374$ & $1.703$ & $3.691$ & $\ \ \,0.182$ & $2.29$ \\
$43$ & $3.73876$ & $1.746$ & $0.795$ & $0.025$ & $0.716$ & $1.2$ & $0.015$ & $0.11$ & $2.001$ & $1.863$ & $3.916$ & $1.156$ & $0.385$ & $1.719$ & $3.639$ & $-0.019$ & $2.99$ \\
$44$ & $3.74344$ & $1.908$ & $0.858$ & $0.023$ & $0.726$ & $1.3$ & $0.024$ & $0.20$ & $2.045$ & $1.918$ & $3.930$ & $1.241$ & $0.414$ & $1.762$ & $3.702$ & $\ \ \,0.253$ & $2.91$ \\
$45$ & $3.76456$ & $1.603$ & $0.862$ & $0.024$ & $0.680$ & $1.2$ & $0.028$ & $0.12$ & $1.952$ & $1.894$ & $3.938$ & $1.259$ & $0.356$ & $1.675$ & $3.696$ & $\ \ \,0.187$ & $3.68$ \\
$46$ & $3.78243$ & $1.757$ & $0.700$ & $0.024$ & $0.681$ & $1.7$ & $0.027$ & $0.05$ & $1.975$ & $1.800$ & $3.940$ & $1.223$ & $0.408$ & $1.753$ & $3.708$ & $\ \ \,0.275$ & $2.51$ \\
$47$ & $3.81793$ & $1.688$ & $0.735$ & $0.026$ & $0.717$ & $1.2$ & $0.013$ & $0.08$ & $1.971$ & $1.798$ & $3.912$ & $1.111$ & $0.339$ & $1.649$ & $3.633$ & $-0.080$ & $3.26$ \\
$48$ & $3.82753$ & $1.726$ & $0.890$ & $0.026$ & $0.735$ & $1.1$ & $0.016$ & $0.11$ & $2.059$ & $1.901$ & $3.911$ & $1.156$ & $0.406$ & $1.750$ & $3.621$ & $-0.075$ & $3.46$ \\
$49$ & $3.82759$ & $1.386$ & $1.012$ & $0.026$ & $0.734$ & $1.3$ & $0.029$ & $0.03$ & $2.029$ & $1.893$ & $3.915$ & $1.168$ & $0.331$ & $1.636$ & $3.646$ & $-0.035$ & $7.11$ \\
$50$ & $3.83383$ & $1.804$ & $0.726$ & $0.021$ & $0.712$ & $1.2$ & $0.025$ & $0.08$ & $2.035$ & $1.899$ & $3.951$ & $1.313$ & $0.372$ & $1.700$ & $3.713$ & $\ \ \,0.265$ & $3.06$ \\
$51$ & $3.84236$ & $1.741$ & $0.701$ & $0.023$ & $0.715$ & $1.2$ & $0.017$ & $0.05$ & $2.004$ & $1.860$ & $3.927$ & $1.201$ & $0.362$ & $1.686$ & $3.657$ & $\ \ \,0.034$ & $3.00$ \\
$52$ & $3.88768$ & $1.877$ & $0.642$ & $0.019$ & $0.739$ & $1.2$ & $0.019$ & $0.05$ & $2.018$ & $1.795$ & $3.936$ & $1.204$ & $0.422$ & $1.772$ & $3.670$ & $\ \ \,0.129$ & $2.53$ \\
$53$ & $3.89491$ & $1.914$ & $0.726$ & $0.025$ & $0.709$ & $1.3$ & $0.022$ & $0.14$ & $2.000$ & $1.845$ & $3.925$ & $1.184$ & $0.425$ & $1.777$ & $3.690$ & $\ \ \,0.212$ & $2.58$ \\
$54$ & $3.89838$ & $1.495$ & $0.909$ & $0.023$ & $0.717$ & $1.1$ & $0.025$ & $0.01$ & $2.031$ & $1.920$ & $3.930$ & $1.242$ & $0.354$ & $1.673$ & $3.644$ & $-0.022$ & $4.81$ \\
$55$ & $3.95017$ & $1.858$ & $0.668$ & $0.026$ & $0.688$ & $1.2$ & $0.024$ & $0.07$ & $2.022$ & $1.901$ & $3.940$ & $1.270$ & $0.394$ & $1.733$ & $3.702$ & $\ \ \,0.238$ & $2.70$ \\
$56$ & $3.96751$ & $2.003$ & $0.683$ & $0.022$ & $0.729$ & $0.9$ & $0.029$ & $0.16$ & $2.019$ & $1.801$ & $3.954$ & $1.279$ & $0.401$ & $1.744$ & $3.724$ & $\ \ \,0.332$ & $2.93$ \\
$57$ & $4.07314$ & $1.895$ & $0.589$ & $0.024$ & $0.711$ & $1.2$ & $0.020$ & $0.02$ & $2.046$ & $1.865$ & $3.936$ & $1.240$ & $0.401$ & $1.743$ & $3.681$ & $\ \ \,0.160$ & $2.49$ \\
$58$ & $4.15402$ & $1.493$ & $0.939$ & $0.025$ & $0.736$ & $0.8$ & $0.023$ & $0.08$ & $1.978$ & $1.883$ & $3.904$ & $1.119$ & $0.356$ & $1.676$ & $3.599$ & $-0.203$ & $5.60$\B\\
\hline
\multicolumn{18}{l}{\textbf{Notes:} $^{\dagger}$ From $\log L/L_{\odot} = 4 \log(T_{\rm eff}/T_{\rm eff \odot}) + 2 \log(R/R_{\odot})$}

\end{tabular}
\end{table*}

\section{Seismic modelling}
\label{sec:SeismicModelling}

\subsection{Fitting the dominant pulsational frequency}
\label{sec:SeismicModelling_non_rot}

\begin{figure}
    \includegraphics[width=\columnwidth,clip]{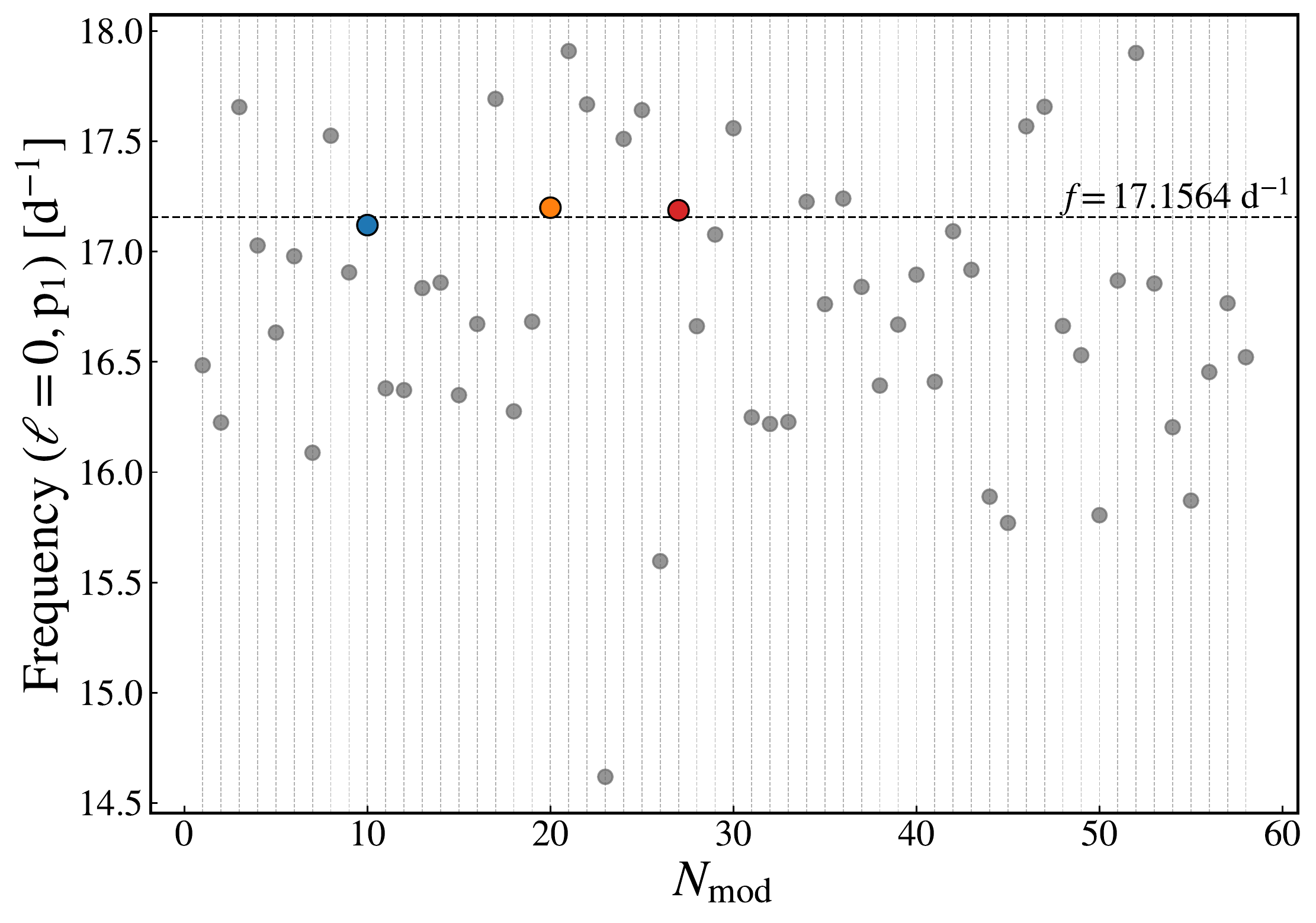}
    \caption{The radial fundamental mode frequencies for the models with initial and final parameters listed in Table\,\ref{tab:fitting_models}.}
    \label{MESA:fitting_models_freqs}
\end{figure}

For the total sample of 58 evolutionary models fitting in masses, radii and orbital period to the observed values, we calculated the radial pulsations for the primary component using the  non-adiabatic linear code of \cite{Dziembowski1977}.
The fundamental mode frequency of these models is in the range of $\sim 14.5-18$\,\cpd, with the closest one to the observed value of $f_1=17.1564$\,\cpd\ 
in the model No.\,27 (see Table\,\ref{tab:fitting_models}), i.e., $f=17.1874$\,\cpd.
This value, however, is far beyond the accepted uncertainty range, i.e. $f_1 \pm \Delta f_{\rm R}$ where $\Delta f_{\rm R} = 0.0046$\,\cpd\ is the Rayleigh resolution.
In Fig.\,\ref{MESA:fitting_models_freqs} we show the fundamental radial mode $(\ell=0, p_1)$ frequency for the considered models as a function of the model number $N_{\rm mod}$.
The colour dots mark the three models ($N_{\rm mod}=10,~20,~27$) with the closest frequency to the observed, and will be discussed later in this Section.

The use of the non-adiabatic pulsation code allows to get the information on pulsational excitation. The mode instability is measured by the parameter $\eta$, that is a normalised work integral computed over one pulsational cycle \citep{Stellingwerf1978}. The value of $\eta$ greater than 0 means that a driving overcomes damping and the pulsation mode is excited (unstable).
For the considered models, in all cases but one, the fundamental radial mode is stable. The instability occurs only in the model No.\,32, however, the frequency of the radial fundamental mode in this model is far from the observed value, i.e. $f~=~16.2188$\,\cpd. For other models, $\eta \in [-0.329, -0.028]$. 
In general, the higher the metallicity, the larger the value of $\eta$. There is also a relation between the value of $T_{\rm eff}$ and $\eta$: the cooler the model, the larger the value of $\eta$.

\begin{figure}
    \includegraphics[width=\columnwidth,clip]{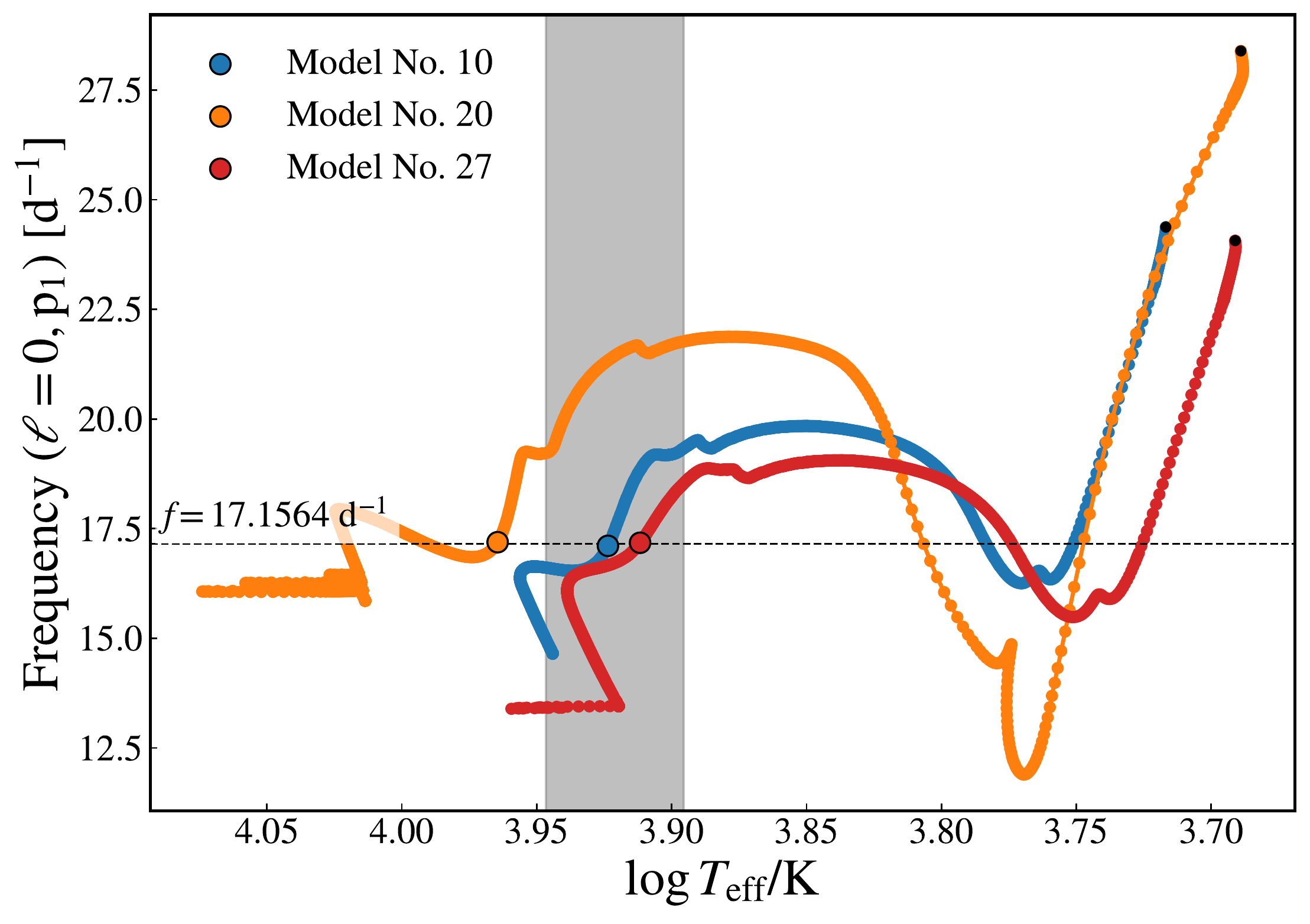}
    \caption{The evolution of the radial fundamental mode frequency. Evolutionary tracks are calculated with the initial parameters for models No.10, 20 and 27 (see Table\,\ref{tab:fitting_models}).}
    \label{MESA:freq_evolution}
\end{figure}

The changes of the mode frequencies during the binary evolution are very different than during the single-star evolution. The dynamic processes that lead to reversing the mass ratio between the components have a significant impact on the evolution of components and on their pulsational properties. Unlike the single-evolution case, where the slow evolution during the main sequence causes slight, monotonic changes in radial mode frequencies, its binary counterpart encounters rapid, non-monotonic changes. From the set of models presented in Table\,\ref{tab:fitting_models}, we selected these that have the fundamental radial mode frequency in the range $17.1 - 17.2$\,\cpd. These are the models No.\,10, 20 and 27. 
In Fig.\,\ref{MESA:freq_evolution} we depicted the evolution of the radial fundamental mode frequency as a function of the effective temperature for evolutionary tracks calculated with the initial parameters of these models.
With black dots we mark the ZAMS models for each of the evolutionary tracks. The vertical shaded area shows the observed effective temperature for the primary in the $3\sigma$ range.
As one can see, each track crosses the $f_1=17.1564$\,\cpd\ line multiple times.
With black circles we marked the models that for the observed orbital period reproduce masses and radii of the components. These models are also marked with colour dots in Fig.\,\ref{MESA:fitting_models_freqs}.

\begin{figure*}
    \includegraphics[width=\textwidth,clip]{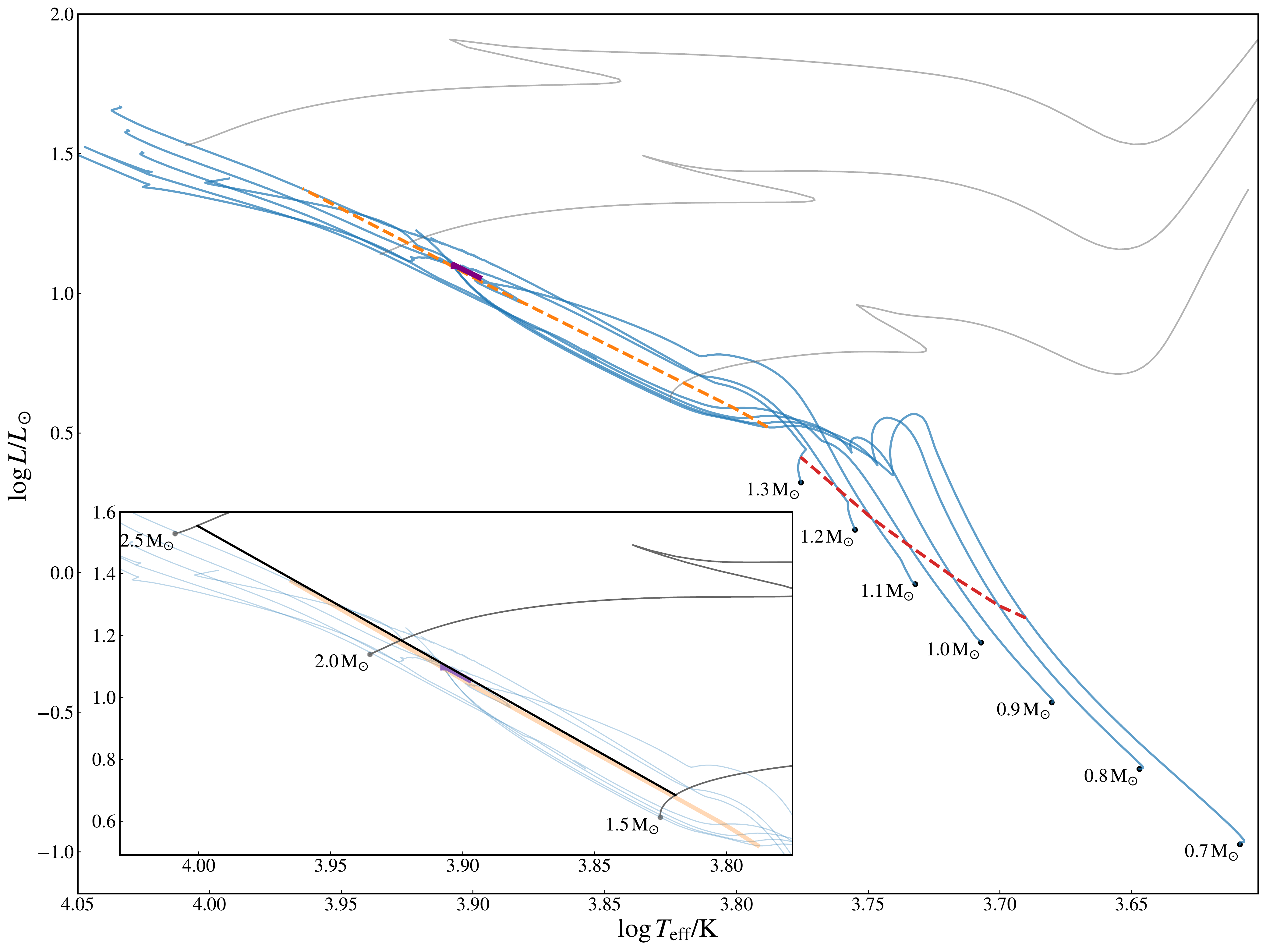}
    \caption{
    The HR diagram with the lines of constant frequency for the radial fundamental mode. With blue lines we plot the binary tracks that were computed for the following parameters: $P_{\rm ini} = 3.7$\,d, $M_{\rm don,ini} = 1.7$\,\Msun, $X_0 = 0.7$, $Z = 0.023$, $f_{\rm ov} = 0.02$, $\alpha_{\rm MLT} = 1.2$ and $\beta = 0.1$, with different initial masses for acceptor. The grey lines show the single-evolution tracks. These tracks were calculated for the same initial hydrogen abundance, metallicity, overshooting and mixing-length parameter as the binary models, and have masses $M=1.5, 2.0$ and $2.5$\,\Msun. Black dots show the location of ZAMS models.
    The lines of constant frequency $f=17.1564$\,\cpd\ of the radial fundamental mode are presented in red, orange and purple. These colors represent the first, second and third crossing of a given frequency, respectively, as demonstrated in Fig.\,\ref{MESA:freq_evolution}. 
    In order to increase clarity between lines of constant frequency in single and binary case, we add the inset in which we plot the single evolutionary tracks against the less transparent binary tracks in the background.
    In the inset, the line of single-evolution constant frequency $f=17.1564$\,\cpd\ of the radial fundamental mode is depicted in black. For clarity, in the main plot this line is not shown.
    }
    \label{MESA:f1_line}
\end{figure*}

In the case of a single-evolution, the frequencies of the radial modes on the evolutionary tracks calculated for different masses determine the lines of constant frequency which are straight lines \citep[e.g.,][]{Pamyatnykh2000,Pamyatnykh2004}. However, a similar behaviour for the binary-evolution has not been tackled in the literature. 
Therefore, in order to demonstrate the differences between the single and the binary-evolution models we built an additional grid of models.
The binary-evolution models were calculated for $P_{\rm ini} = 3.7$\,d, $M_{\rm don,ini} = 1.7$\,\Msun, $X_0 = 0.7$, $Z = 0.023$, $f_{\rm ov} = 0.02$, $\alpha_{\rm MLT} = 1.2$ and $\beta = 0.1$, assuming various initial masses for acceptor, i.e., $M_{\rm acc,ini} \in [0.7, 1.3]$\,\Msun\ with the step of $0.1$\,\Msun. 
For a comparison, we calculated single-evolution models with the same initial hydrogen abundance, metallicity, overshooting and mixing-length parameter, for masses $M=1.5, 2.0$ and $2.5$\,\Msun.
In Fig.\,\ref{MESA:f1_line}, we show the HR diagram with evolutionary tracks for single (grey lines) and binary models (blue lines) for the acceptor.
The line of constant frequency $f=17.1564$\,\cpd\ of the radial fundamental mode in the case of single-evolution models is depicted in black (for clarity we show this line only in the inset).
In red, orange and purple we marked the lines of constant frequency $f=17.1564$\,\cpd\ in the case of binary-evolution models. These colors represent the first, second and the third crossing of a given frequency, respectively, as demonstrated in Fig.\,\ref{MESA:freq_evolution}. 
The frequency $f=17.1564$\,\cpd\ is not reproduced the same number of times during the evolution for each mass.

\begin{figure*}
    \centering
    \begin{tabular}{cc}
    \includegraphics[width=\columnwidth,clip]{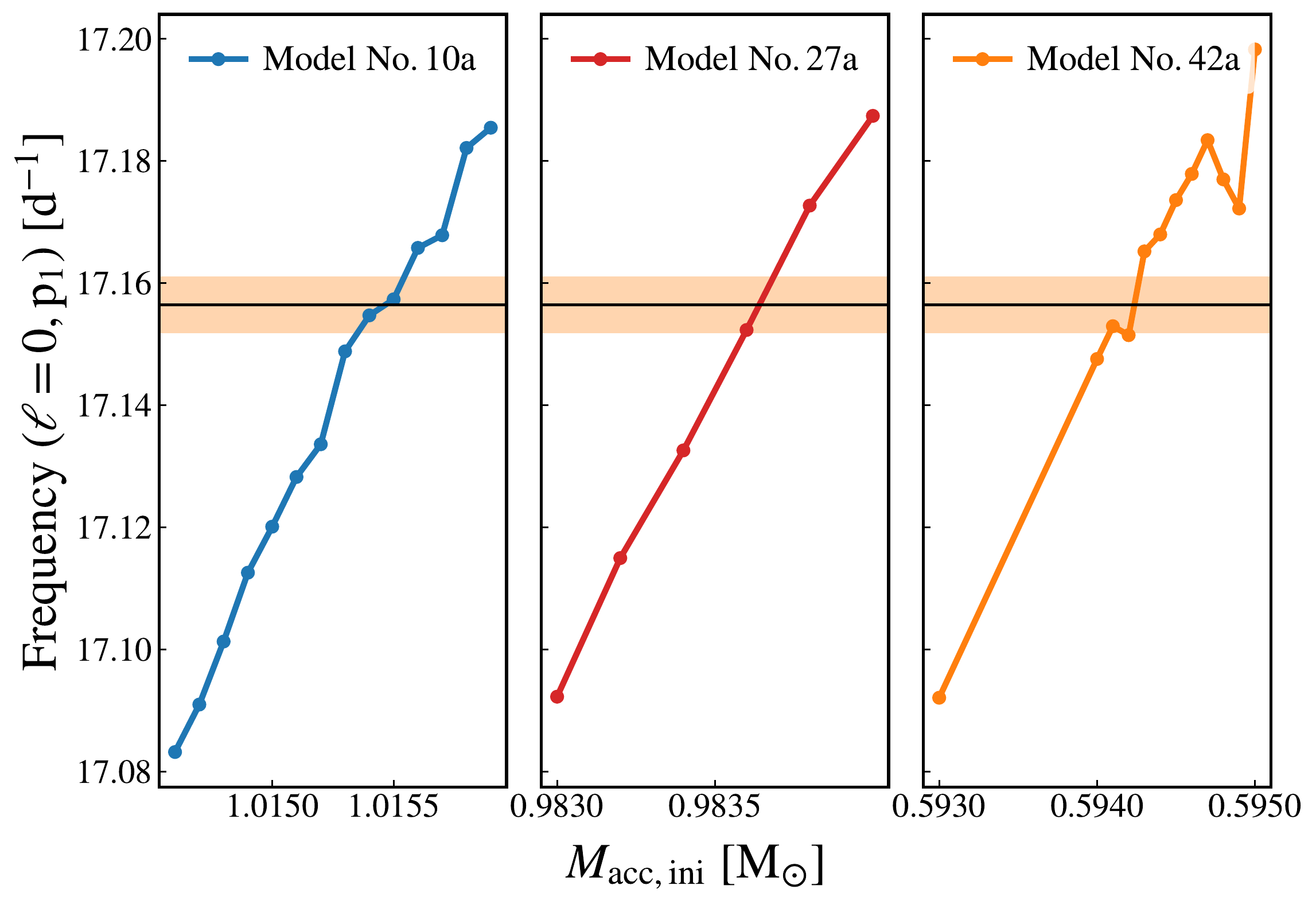} &
    \includegraphics[width=\columnwidth,clip]{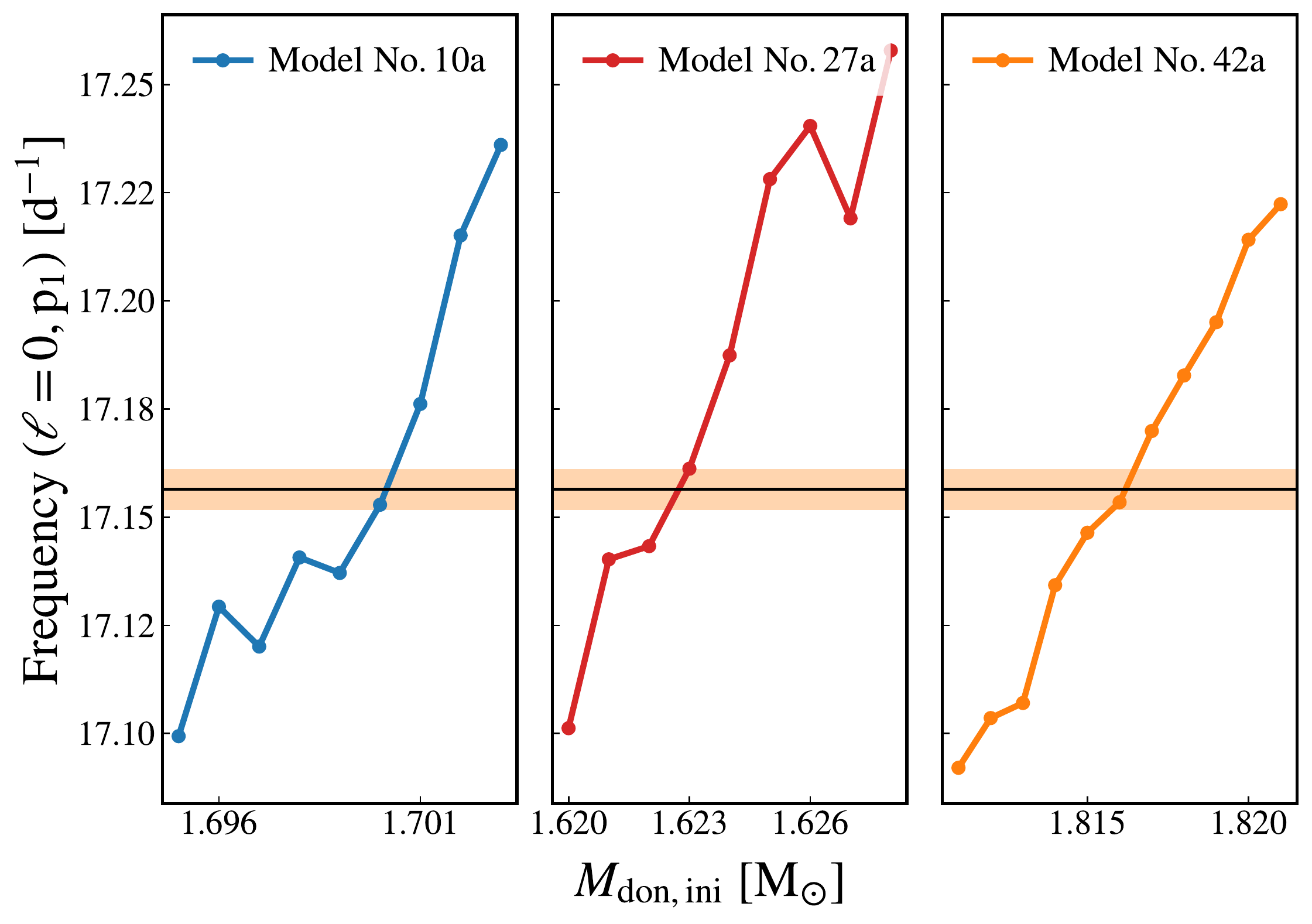} \\
    \includegraphics[width=\columnwidth,clip]{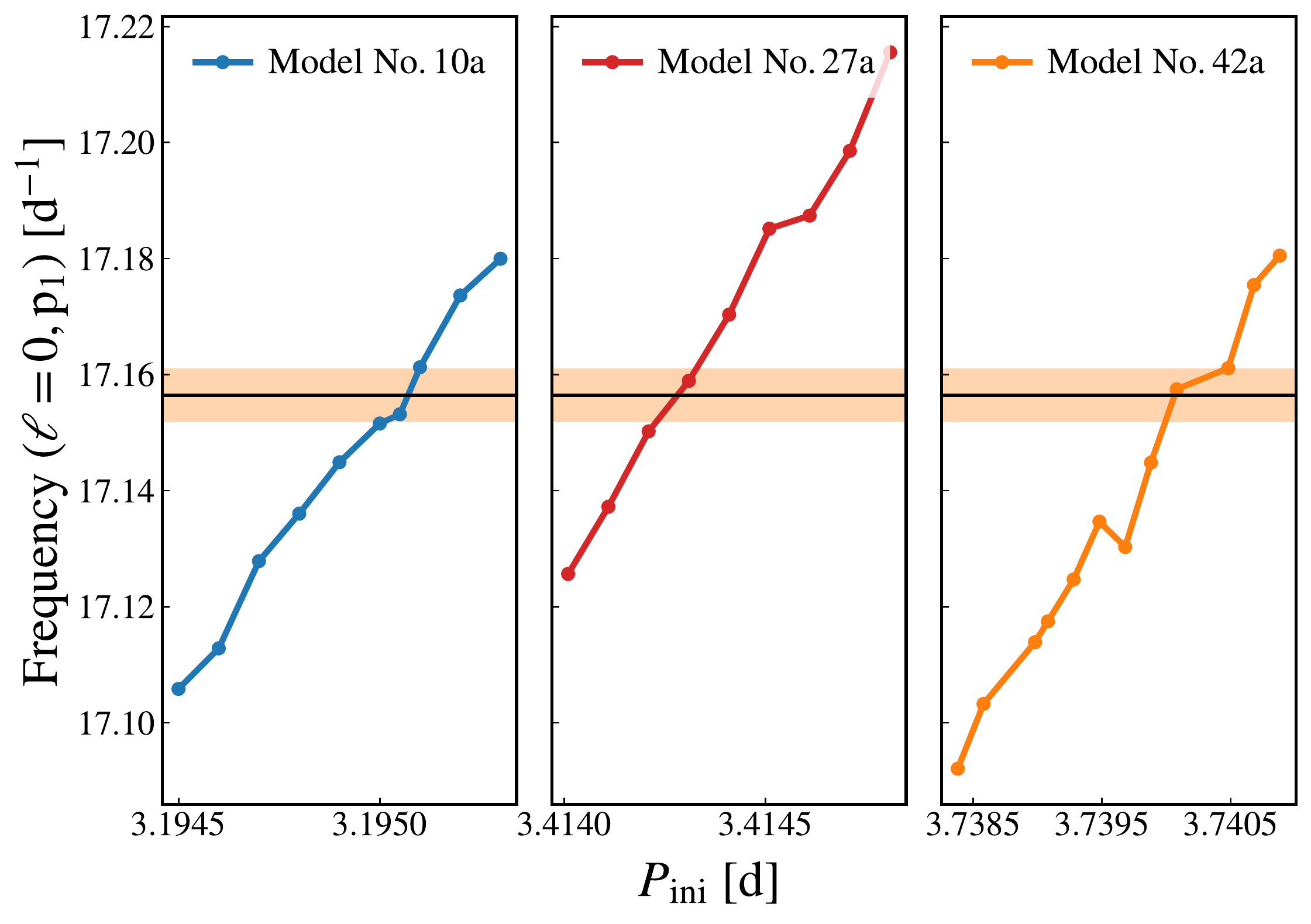} &
    \includegraphics[width=\columnwidth,clip]{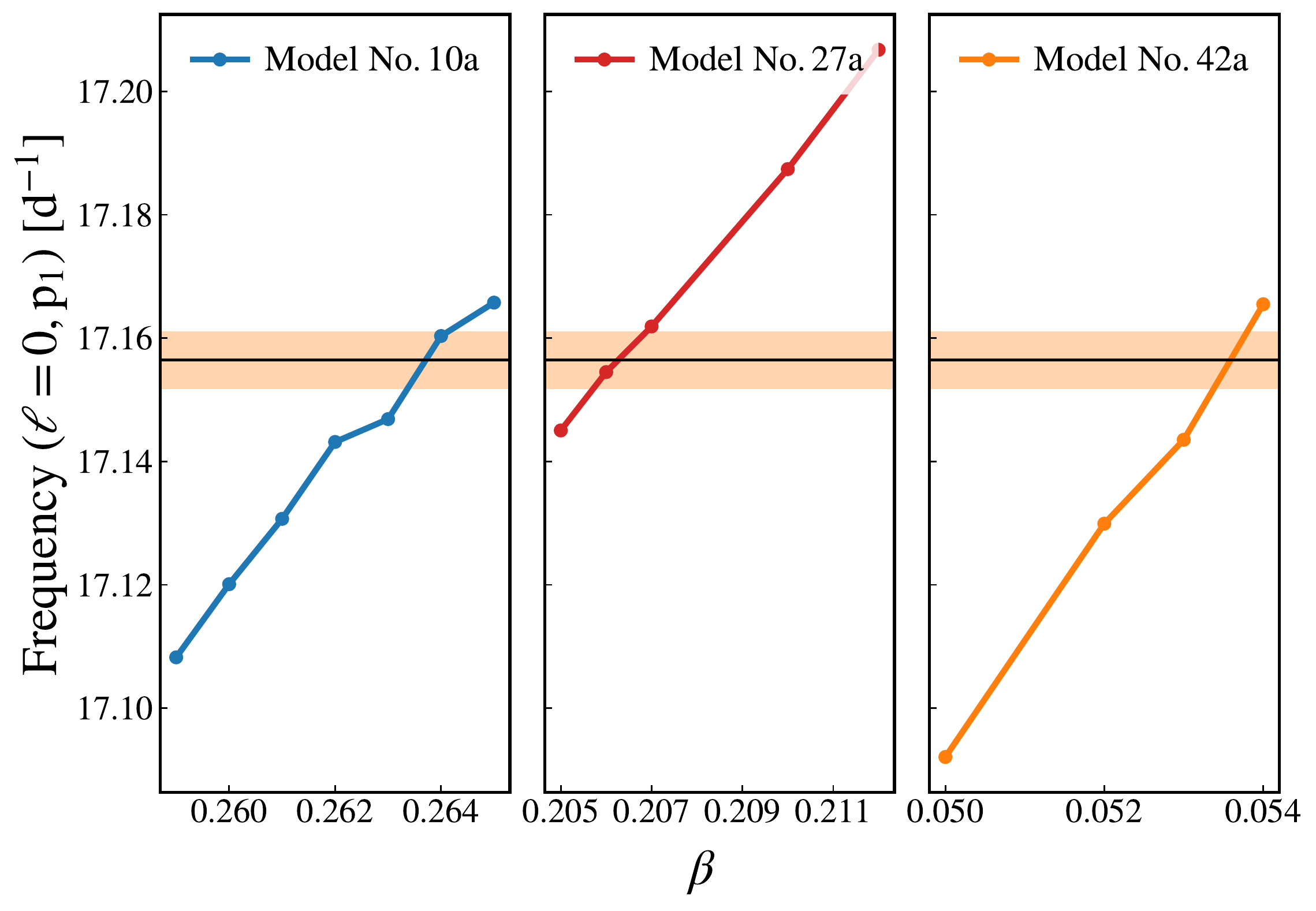}
    \end{tabular}
    
    \caption{Dependencies of the theoretical frequency on the adopted parameters for models No. 10, 20 and 27. With the horizontal solid lines and the shaded areas we mark the observed value of the dominant frequency with the allowed uncertainty range, $f_1 \pm \Delta f_{\rm R}$.}
    \label{MESA:freq_dependency}
\end{figure*}

Small changes in the initial parameters that are used to find models can have a noticeable impact on the structure of resulting stars. For example, a small change in the mass transfer rate can cause the final mass and radius of the star to be sightly different. By that, the dynamical time scale, that in approximation characterizes the pulsational period of the fundamental radial mode, can change. This results in a slight increase or decrease of the frequency.
Following that approach, we proceeded with models that have the closest fundamental mode frequency to the observed value.
The models with numbers 10, 20 and 27 have the closest values of $f(\ell=0, \rm p_1)$ but none of them fit in the effective temperatures for both components simultaneously (see Table\,\ref{tab:PHOEBE_paramterers}). However, the models 10 and 27 have the values of $T_{\rm eff}$ for the primary component in the allowed range.
Moreover, the model 42 reproduces $T_{\rm eff}$ for both the acceptor and donor, and has the radial fundamental mode frequency $f=17.092085$\,\cpd. Therefore, we used this model along with the models No.\,10 and 27 in our further fine-tuning seismic modelling. For these three models we calculated additional small evolutionary grids, changing the initial values of $P_{\rm ini}$, $M_{\rm acc,ini}$, $M_{\rm don,ini}$, $Z$, $f_{\rm ov}$ and $\beta$, in the closest vicinity of their parameters listed in Table\,\ref{tab:fitting_models}. 
We shall refer to these fine-tuned models by using an "a" subscript, i.e., 10a, 27a and 42a.
We found that such parameter changes cause, in general, monotonic changes in the frequency of the fundamental radial mode.
The growth of $P_{\rm ini}$, $M_{\rm don,ini}$, $M_{\rm acc,ini}$, $\beta$, and the decrease of $Z$, $f_{\rm ov}$ parameters increase the fundamental mode frequency, so it is straightforward to extract the parameters that reproduce the observed value of $f_1$ using the linear regression. We plot the discussed dependencies of the frequency $f(\ell=0, \rm p_1)$ on the considered parameters in Fig.\,\ref{MESA:freq_dependency}. With the shaded area we mark the allowed uncertainty range of $f_1$ value, i.e. $f_1 \pm \Delta f_{\rm R}$. We also note that changing $Z$ or $f_{\rm ov}$ has a big effect on the frequency. Even slight changes of their value, with an order of $10^{-6}$, dramatically impact the $f_1$ value, therefore, they are very strongly constrained by the fitting of the pulsational frequency.
We summarize all the solutions in Table\,\ref{tab:final_paramterers}. In this Table we also provide the values of $\eta(\ell=0, \rm p_1)$ for the final models.

\begin{table}
    \centering
    \footnotesize
    \caption{The parameters of the models that reproduce the observed value of $f_1=17.1564$\,\cpd\ as the radial fundamental mode.}
    \label{tab:final_paramterers}
    \begin{tabular}{llll}
    
        \hline
        \hline
        Parameter & No. 10a & No. 27a & No. 42a \T\B \\
        \hline
        $P_{\rm ini}\,\rm[d]$       & $3.19505(4)$ & $3.41428(5)$ & $3.74024(24)$ \T\\
        $M_{\rm don,ini}\,\rm[M_{\odot}]$  & $1.7001(2)$ & $1.62277(6)$ & $1.81612(36)$ \\
        $M_{\rm acc,ini}\,\rm[M_{\odot}]$   & $1.01545(5)$ & $0.9836(5)$ & $0.59418(10)$  \\
        $R_{\rm don}$\,[\Rsun]   & $1.715$ & $1.700$ & $1.703$ \\
        $R_{\rm acc}$\,[\Rsun]   & $1.849$ & $1.837$ & $1.816$ \\
        $\log T_{\rm eff}^{\rm don}$ & $3.642$ & $3.619$ & $3.691$ \\
        $\log T_{\rm eff}^{\rm acc}$ & $3.923$ & $3.912$ & $3.942$ \\
        $X_0$               & $0.734$ & $0.726$ & $0.682$ \\
        $Z$                 & $0.019$ & $0.023$ & $0.022$ \\
        $f_{\rm ov}$        & $0.011$ & $0.011$ & $0.018$ \\
        $\alpha_{\rm MLT}$  & $1.4$ & $1.2$ & $1.2$ \\
        $\beta$             & $0.2638(4)$ & $0.2063(6)$ & $0.0537(3)$ \\
        $\eta(\ell=0,\rm p_1)$  & $-0.12$ & $-0.06$ & $-0.10$ \\
        Age [Gyr]  & $2.85$ & $3.38$ &  $2.92$ \B\\

        \hline

    \end{tabular}
\end{table}

\subsection{Estimating the effects of rotation}
\label{sec:SeismicModelling_rot}

The main drawback of our seismic modelling in the previous subsection is the complete neglect of the rotation effects, both on the equilibrium models and on pulsational  frequencies.
However, in the case of radial modes, it is relatively easy to estimate the effects of rotation on the frequency value.
Let us recall the third order expression for a rotationally split frequency  \citep[e.g.][]{Gough1990,Soufi1998,Goupil2000,Pamyatnykh2003}:
\begin{align}
    \omega_{n\ell m} & = \omega_0 + m\left( 1- C_{n\ell} \right) \Omega + \frac{\Omega^2}{\omega_0}\left( D_0 + m^2D_1 \right) + m\frac{\Omega^3}{\omega_0^2}T. \label{eqn:rot_splitting}
\end{align}
In this formula, $\omega_0$ includes the effects of rotation in the equilibrium model.
Here, $\Omega$ is the angular rotational frequency and $C_{n\ell}$ is the Ledoux constant, which depends on the radial order $n$ and the degree $\ell$ of a given mode. 
This constant determines the equidistant splitting of the frequency in the limit of slow rotation. 

In the case of AB\,Cas, the rotational velocity of the main component is about $V_{\rm rot} = 70$\,km/s. To estimate the frequency difference of the fundamental radial mode between the non-rotating and rotating equilibrium model, we computed the single-evolution models with $M=2.01$\Msun, assuming  $V_{\rm rot}=0$ and $V_{\rm rot}=70$\,km/s.
For the mode $\ell=0,\rm~p_1$ with the frequency around the observed value, the frequency difference between
non-rotating and rotating models amounts to about $0.1$\,\cpd.

According to Eq.\,\ref{eqn:rot_splitting}, for radial modes only the term $D_0$ remains and is equal to 4/3 independently of the radial order, $n$ \citep[e.g.,][]{Kjeldsen1998}. Thus, we have
\begin{align}
\omega=\omega_0 + \frac{4}{3} \frac{\Omega^2}{\omega_0} \label{eqn:rot_splitting_m0}
\end{align}
or
\begin{align}
f=f_0 + \frac{4}{3} \frac{f_{\rm rot}^2}{f_0}. \label{eqn:rot_splitting_m0}
\end{align}
The above expression was found first by \cite{Simon1969}.

Taking the observed value of the dominant frequency, i.e., $f=17.1564$\,\cpd,
one gets $f_0=17.1147$\,\cpd\ from the formula Eq.\,\ref{eqn:rot_splitting_m0}. Therefore, the second order effect of rotation 
is about $\Delta f=0.04$\,\cpd.

Another  important effect of rotation on radial mode frequencies is rotational mode coupling \citep{Soufi1998}.
It occurs if the modes with harmonic degrees differing by two have very close frequencies,
i.e., their difference is of the order of a rotational frequency. Thus the conditions are:
$\omega_j-\omega_k\approx\Omega, ~~\ell_j=\ell_k+2,~~m_j=m_k$.
Taking the models fitting the frequency $f_1$, we found that the modes $\ell,\rm~p_1$ and $\ell=2,\rm~g_2$ have the closest frequencies.
The frequency difference between these two modes was about $0.2$\,\cpd\ that is much smaller than the rotational frequency ($f_{\rm rot}=0.73159$\,\cpd).
The rotational mode coupling between these two modes gives the frequency shift of the mode $\ell=0,\rm ~p_1$ of about $0.03$\,\cpd.
The subsequent effects  of rotation on pulsational frequencies  are presented, for example, by \cite{Pamyatnykh2003}.

Thus, the total uncertainty in the pulsational frequency resulting from the fact that we did not take into account the effects of rotation is about $\Delta f = 0.17$\,\cpd. 
We adopted this value and accepted all models with the frequency of the radial fundamental mode in the range $f = 17.1564 \pm 0.17$\,\cpd.
These are models number 4, 6, 10, 20, 27, 29, 34, 36 and 42 (see Tab.\,\ref{tab:fitting_models} and Fig.\,\ref{MESA:fitting_models_freqs}). All of these models reproduce the $f(\ell = 0, \rm p_1)$ in the allowed range. However, only models 6, 10, 27, 29, 34, 42 reproduce the value of effective temperature for the primary star, with models 6 and 42 reproducing $\log T_{\rm eff}$ also for the secondary component. The parameters of the models 6, 10, 27, 29, 34, 42 are summarised in Tab.\,\ref{tab:final_paramterers_rot}.

\begin{table*}
    \centering
    \caption{The parameters of the models that reproduce the observed value of dominant frequency $f_1=17.1564 \pm 0.17$\,\cpd\ as the radial fundamental mode, while taking the rotation effects into account.}
    \label{tab:final_paramterers_rot}
    \begin{tabular}{lrrrrrr}
    
        \hline
        \hline
        Parameter                           & No. 6   & No. 10  & No. 27 & No. 29 & No. 34 & No. 42 \T\B \\
        \hline
        $P_{\rm ini}\,\rm[d]$             & $3.04530$&$3.19466$&$3.41461$ &$3.43473$&$3.55946$& $3.73838$  \T\\
        $M_{\rm don,ini}\,\rm[M_{\odot}]$   & $1.785$ & $1.667$ & $1.624$ & $1.769$ & $1.651$ & $1.811$ \\
        $M_{\rm acc,ini}\,\rm[M_{\odot}]$   & $0.962$ & $1.015$ & $0.984$ & $0.732$ & $0.799$ & $0.593$ \\
        $R_{\rm don}$\,[\Rsun]              & $1.731$ & $1.715$ & $1.700$ & $1.730$ & $1.760$ & $1.703$ \\
        $R_{\rm acc}$\,[\Rsun]              & $1.854$ & $1.849$ & $1.837$ & $1.867$ & $1.841$ & $1.816$ \\
        $\log T_{\rm eff}^{\rm don}$        & $3.679$ & $3.642$ & $3.619$ & $3.643$ & $3.645$ & $3.691$ \\
        $\log T_{\rm eff}^{\rm acc}$        & $3.938$ & $3.923$ & $3.912$ & $3.938$ & $3.936$ & $3.942$ \\
        $X_0$                               & $0.734$ & $0.734$ & $0.726$ & $0.707$ & $0.696$ & $0.682$ \\
        $Z$                                 & $0.015$ & $0.019$ & $0.023$ & $0.013$ & $0.021$ & $0.022$ \\
        $f_{\rm ov}$                        & $0.015$ & $0.011$ & $0.011$ & $0.010$ & $0.017$ & $0.018$ \\
        $\alpha_{\rm MLT}$                  & $1.5$   & $1.4$   & $1.2$   & $1.3$   & $1.2$   & $1.3$   \\
        $\beta$                             & $0.28$  & $0.26$  & $0.21$  & $0.05$  & $0.04$  & $0.05$  \\
        $\eta(\ell=0,\rm p_1)$              & $-0.20$ & $-0.12$ & $-0.06$ & $-0.18$ & $-0.17$ & $-0.10$ \\
        Age [Gyr]                           & $2.46$  & $2.85$  & $3.38$  & $2.33$  & $2.83$  & $2.29$ \B\\

        \hline

    \end{tabular}
\end{table*}

\section{Discussion and conclusions}
\label{sec:conclusions}

We performed comprehensive studies of the AB\,Cas binary system containing a pulsating star of \dsct\ type. First, for the Str\"omgren data gathered from the literature and the \textit{TESS} observations from three sectors, we modelled the eclipsing light curve of the system using the \PHOEBE\ code. We found that the donor star (the former primary), while filling its Roche lobe, is still transferring mass to the acceptor. The deformation of the donor and the proximity to its companion cause large inequalities in the effective temperature distribution on its surface. Due to heating from the primary star, its Roche tail is hotter of around 500\,K than the hemisphere that is facing outwards the system. Moreover, some trends in the light curve residua can suggest the presence of an accretion disk in the system and associated with the disk hot spots on the primary star.

After subtraction of the model eclipsing  light curve, we looked for variability in the residua. By applying the Fourier analysis and standard pre-whitening procedure we found 112 significant frequencies with the signal-to-noise ratio $S/N > 4$. Amongst them, we found combination frequencies with the orbital frequency. 
In total, 17 frequencies seem to be independent. The most prominent signal, $f_1=17.1564$\,\cpd, is the radial fundamental mode which was already reporetd in the literature. We also confirmed the presence of two other peaks in the data, that were mentioned by \cite{Rodriguez2004}, i.e. $f=18.25$\,\cpd\ and $f=34.64$\,\cpd.

Using the times of the light minima gathered in the literature, we re-determined the change of the orbital period ratio. By fitting the second order polynomial to the $(O-C)$ data, we concluded that the system experiences an increase of the orbital period with a rate of 0.03 s/yr. This result is consistent with the values reported in \cite{Soydugan2003}, but slightly lower comparing to \cite{Abedi2007}. Moreover, the $(O-C)$ data can also suggest the presence of third light in the system, with the orbital period higher than 25 years.

In order to establish the current evolutionary state of AB\,Cas, we computed the binary-evolution of the system using the \MESAbinary\ code. We found over a dozen of models that for an observed value of the orbital period reproduce the masses, the radii and the effective temperatures of the components within 3$\sigma$ errors. These models indicate the age of the system to be between $1.86 - 4.26$\,Gyr.
The mass transfer event started when the central hydrogen abundance of the donor has dropped below the value of $X_c \approx 0.06 - 0.10$ indicating the case A of MT. This value points that the donor was about to transit to the overall-contraction phase and to reconstruct its internal structure. On the other hand, the current evolutionary phase of the system suggests that the primary star, after accreting most of the transferred mass, is evolving on the main sequence, with the central hydrogen abundance of about $X_c \approx 0.40 - 0.56$. The secondary star is just after the core conversion to fully helium, and is currently burning hydrogen in the shell. In the future, it will lead the secondary to rapid H-shell flashes, during which the hydrogen shell will be substantially reduced leading the star to the cooling helium white dwarf stage.
We have also determined the initial parameters of the system and its components that can help to explain the evolution history of AB\,Cas.
These parameters are: $P_{\rm ini} \in[3.04530, 4.07314]$\,d, $M_{\rm don,ini} \in [1.573, 2.042]$\,\Msun, $M_{\rm acc,ini} \in [0.589, 1.033]$\,\Msun, $Z \in [0.015, 0.025]$, $f_{\rm ov} \in [0.009, 0.028]$, $\alpha_{\rm MLT} \in [1.1, 1.6]$ and $\beta \in [0.05, 0.28]$.

In the next step, we constructed the pulsational models that reproduce the dominant frequency as the radial fundamental mode. With the fine-tuning procedure,
we found the parameters of the model that reproduce $f_1$ with the Rayleigh resolution.
For the first time we showed how the frequency of the fundamental radial mode changes during the binary evolution.
By changing one of the parameters of the model, while keeping the others constant, the change of the radial fundamental mode frequency is quasi-monotonic. Such behaviour is typical for all of the considered parameters, as $P_{\rm ini}$, $M_{\rm don,ini}$, $M_{\rm acc,ini}$, $X_0$, $Z$, $f_{\rm ov}$, $\alpha_{\rm MLT}$ and $\beta$. We also showed that, contrary to single evolutionary modelling, the constant frequency line for the binary case is not just one straight line in the HR diagram. The evolutionary changes of the radial mode frequencies
are not monotonic and their character strongly depends on the model parameters. Such behaviour produces multiple lines of the constant frequency in the HR diagram.

Our seismic models give the following constraints on the parameters used in the binary-evolution modelling:
$P_{\rm ini}~\in~[3.195, 3.740]$\,d, $M_{\rm don,ini}~\in~[1.623, 1.816]$\,\Msun, $M_{\rm acc,ini} \in [0.594, 1.015]$\,\Msun, $X_0 \in [0.68, 0.73]$, $Z \in [0.019, 0.023]$, $f_{\rm ov} \in [0.011, 0.018]$, $\alpha_{\rm MLT} \in [1.2, 1.4]$ and $\beta \in [0.05, 0.26]$. The system is about 3 Gyr old, and consists of components with the radii: $R_{\rm acc} \in [1.82, 1.85]$\,\Rsun, $R_{\rm don} \in [1.70, 1.72]$\,\Rsun\ and with effective temperatures: $\log T_{\rm eff}^{\rm acc}\in [3.91,3.94]$ and $\log T_{\rm eff}^{\rm don}\in[3.62,3.69]$.

Given the importance of the effects of rotation, both, on the equilibrium model as well as on pulsational frequencies, we roughly estimated the frequency shifts caused by these effects. For the radial fundamental mode, we found the uncertainty of about $\Delta f = 0.17$\,\cpd\ which results from the effects of rotation on the equilibrium model and from the third order effects of rotation on pulsational frequencies. With this frequency uncertainty, the models that reproduce the dominant frequency have the following parameters: $P_{\rm ini} \in[3.04530, 3.73838]$\,d, $M_{\rm don,ini} \in [1.624, 1.811]$\,\Msun, $M_{\rm acc,ini} \in [0.593, 1.015]$\,\Msun, $Z \in [0.015, 0.023]$, $f_{\rm ov} \in [0.010, 0.018]$, $\alpha_{\rm MLT} \in [1.2, 1.5]$ and $\beta \in [0.04, 0.28]$.

Similarly to the first paper in the series of "The eclipsing binary systems with $\delta$\,Scuti component", which is devoted to KIC\,10661783, we showed that close systems should be modelled as binary and not as single stars. Only the multifaceted studies based on the binary-evolution models lead to the reliable reconstruction of the evolutionary past of such systems and allow to draw more general conclusions on their evolution. 
In order to further develop the binary asteroseismology, we plan to study more systems of this type with a pulsating \dsct\ component in this comprehensive way.

\section*{Acknowledgements}
The authors are grateful to the referee for constructive feedback that helped to improve this paper.
This work was financially supported by the Polish National Science Centre grant 2018/29/B/ST9/02803. \\
PKS have been supported by the Polish National Science Center grant no.~2019/35/N/ST9/03805. \\
Calculations have been carried out using resources provided by Wroc\l aw Centre for
Networking and Supercomputing (http://wcss.pl), grant no. 265. \\
This paper includes data collected by the \textit{TESS} mission. Funding for the \textit{TESS} mission is provided by the NASA Explorer Program. Funding for the \textit{TESS} Asteroseismic Science Operations Centre is provided by the Danish National Research Foundation (Grant agreement no.: DNRF106), ESA PRODEX (PEA 4000119301) and Stellar Astrophysics Centre (SAC) at Aarhus University. We thank the \textit{TESS} team and staff and TASC/TASOC for their support of the present work.

\section*{Data Availability}

The target pixel files were downloaded from the public data archive at MAST. The light curves will be shared upon a reasonable request. The full list of frequencies is available as a supplementary material to this paper. We make all inlists needed to recreate our \texttt{MESA-binary} results publicly available at Zenodo. These can be downloaded at \url{https://doi.org/10.5281/zenodo.5806513}.

\section*{Software}

-- \texttt{MESA} \citep{Paxton2011,Paxton2013,Paxton2015,Paxton2018,Paxton2019} \\
-- \texttt{LIGHTKURVE} \citep{lightkurve2018} \\
-- \texttt{PHOEBE\,2} \citep{Prsa2005,Prsa2016,Horvat2018,Jones2020,Conroy2020} \\
-- \texttt{Python SciPy} \citep{2020SciPy-NMeth} \\

\section*{Orcid IDs}

\flushleft
A. Miszuda\orcidAM\ \url{https://orcid.org/0000-0002-9382-2542} \\
P. A. Ko\l aczek-Szyma\'nski\orcidPKS\  \url{https://orcid.org/0000-0003-2244-1512} \\
W. Szewczuk\orcidWS\ \url{https://orcid.org/0000-0002-2393-8427} \\
J. Daszy\'nska-Daszkiewicz\orcidJDD\  \url{https://orcid.org/0000-0001-9704-6408} \\




\bibliographystyle{mnras}
\interlinepenalty=10000
\bibliography{miszuda} 



\appendix

\section{List of observed frequencies}
\label{sec:appendix}

\begin{table*}
    \caption{Complete list of observed frequencies from \textit{TESS} data. The columns contain ID number, frequency, amplitude and phase with their errors respectively, S/N ratio and remarks.}
    \label{tab:freqs}
    \begin{tabular}{lcccccccc}
        \hline
        ID & Frequency $f[\rm d^{-1}]$ & $\sigma_f [\rm d^{-1}]$ & Amplitude $A[\rm ppt]$ & $\sigma_A [\rm ppt]$ & Phase $\Phi [\rm rad]$ & $\sigma_{\Phi} [\rm rad]$ & S/N & Remarks \\
        \hline
        \hline
$f_{ 1 } $ & $  17.1564301120 $ & $   0.0000031811 $ & $  12.3018177011 $ & $   0.0119530009 $ & $   0.1399924969 $ & $   0.0001547013 $ & $  18.15 $ & \\
$f_{ 2 } $ & $  18.3246610620 $ & $   0.0000295814 $ & $   1.2043310480 $ & $   0.0124814052 $ & $   0.0494667850 $ & $   0.0016489619 $ & $   8.03 $ & \\
$f_{ 3 } $ & $  18.2436052200 $ & $   0.0000288402 $ & $   1.1995840861 $ & $   0.0124170335 $ & $   0.2639422414 $ & $   0.0016482140 $ & $   9.87 $ & \\
$f_{ 4 } $ & $  33.0037939610 $ & $   0.0000368047 $ & $   1.0214014573 $ & $   0.0132823382 $ & $   0.4812487113 $ & $   0.0020694699 $ & $  15.34 $ & \\
$f_{ 5 } $ & $  16.7803635080 $ & $   0.0000381662 $ & $   0.9857430791 $ & $   0.0124841973 $ & $   0.5821357341 $ & $   0.0020144961 $ & $   8.02 $ & $ f_{    3}    -2 f_{\rm orb} $ \\
$f_{ 6 } $ & $  17.8879772630 $ & $   0.0000407971 $ & $   0.8288333548 $ & $   0.0119359337 $ & $   0.8949134450 $ & $   0.0022919483 $ & $   8.66 $ & $ f_{    1} +  f_{\rm orb} $ \\
$f_{ 7 } $ & $   0.0658648990 $ & $   0.0000457506 $ & $   0.7597620598 $ & $   0.0115947367 $ & $   0.0803710899 $ & $   0.0025611524 $ & $   5.00 $ & \\
$f_{ 8 } $ & $  16.4248795750 $ & $   0.0000419945 $ & $   0.7844217330 $ & $   0.0119289462 $ & $   0.4294851903 $ & $   0.0024206607 $ & $   7.64 $ & $ f_{    1} - f_{\rm orb} $ \\
$f_{ 9 } $ & $  14.2300583470 $ & $   0.0000437135 $ & $   0.7449765337 $ & $   0.0118992392 $ & $   0.0686116847 $ & $   0.0025417988 $ & $  14.65 $ & $ f_{    1}    -4 f_{\rm orb} $ \\
$f_{ 10 } $ & $  19.3512787560 $ & $   0.0000427350 $ & $   0.7364630995 $ & $   0.0119063577 $ & $   0.9832661514 $ & $   0.0025731896 $ & $  11.79 $ & $ f_{    1} +     3 f_{\rm orb} $ \\
$f_{ 11 } $ & $  20.0827785640 $ & $   0.0000419057 $ & $   0.7544372962 $ & $   0.0119124824 $ & $   0.2960063656 $ & $   0.0025119826 $ & $  10.72 $ & $ f_{    1} +     4 f_{\rm orb} $ \\
$f_{ 12 } $ & $  14.9615569410 $ & $   0.0000424640 $ & $   0.7190355139 $ & $   0.0119011192 $ & $   0.3949543359 $ & $   0.0026338978 $ & $  13.74 $ & $ f_{    1}    -3 f_{\rm orb} $ \\
$f_{ 13 } $ & $  34.3127959770 $ & $   0.0000452491 $ & $   0.6160900186 $ & $   0.0119792050 $ & $   0.2523592362 $ & $   0.0030945271 $ & $   9.60 $ & $    2 f_{    1} $ \\
$f_{ 14 } $ & $  18.6198968370 $ & $   0.0000460742 $ & $   0.6382072353 $ & $   0.0119219971 $ & $   0.3630140003 $ & $   0.0029730339 $ & $   8.75 $ & $ f_{    1} +     2 f_{\rm orb} $ \\
$f_{ 15 } $ & $  15.6931563790 $ & $   0.0000464663 $ & $   0.6160867336 $ & $   0.0119113862 $ & $   0.4369852201 $ & $   0.0030757869 $ & $  11.04 $ & $ f_{    1}    -2 f_{\rm orb} $ \\
$f_{ 16 } $ & $  13.4985146600 $ & $   0.0000507104 $ & $   0.5659829875 $ & $   0.0119207733 $ & $   0.8435983785 $ & $   0.0033529198 $ & $  13.80 $ & $ f_{    1}    -5 f_{\rm orb} $ \\
$f_{ 17 } $ & $  34.4669111270 $ & $   0.0000503062 $ & $   0.6280510176 $ & $   0.0136496617 $ & $   0.8992850817 $ & $   0.0034596348 $ & $  10.67 $ & $ f_{    4} +     2 f_{\rm orb} $ \\
$f_{ 18 } $ & $  20.8143807620 $ & $   0.0000507723 $ & $   0.5530704027 $ & $   0.0118996275 $ & $   0.4379201983 $ & $   0.0034238254 $ & $  11.61 $ & $ f_{    1} +     5 f_{\rm orb} $ \\
$f_{ 19 } $ & $  16.6783723280 $ & $   0.0000513883 $ & $   0.5132379431 $ & $   0.0119803611 $ & $   0.6007652972 $ & $   0.0037155886 $ & $   8.17 $ & \\
$f_{ 20 } $ & $  21.5459766490 $ & $   0.0000520832 $ & $   0.5034333028 $ & $   0.0119132223 $ & $   0.5694260188 $ & $   0.0037676719 $ & $   9.75 $ & $ f_{    1} +     6 f_{\rm orb} $ \\
$f_{ 21 } $ & $  12.7669556060 $ & $   0.0000558616 $ & $   0.4860387310 $ & $   0.0118960220 $ & $   0.6230312797 $ & $   0.0038968782 $ & $  13.67 $ & $ f_{    1}    -6 f_{\rm orb} $ \\
$f_{ 22 } $ & $  18.2582527040 $ & $   0.0000568817 $ & $   0.5058118891 $ & $   0.0125467797 $ & $   0.0536126562 $ & $   0.0039478149 $ & $   9.15 $ & $ f_{    2} - f_{    7} $ \\
$f_{ 23 } $ & $  32.2807524780 $ & $   0.0000627897 $ & $   0.3585322265 $ & $   0.0122146816 $ & $   0.5933364862 $ & $   0.0054216217 $ & $   9.43 $ & \\
$f_{ 24 } $ & $  28.2896774740 $ & $   0.0000651472 $ & $   0.3983375451 $ & $   0.0119055303 $ & $   0.0563191943 $ & $   0.0047559621 $ & $  12.00 $ & \\
$f_{ 25 } $ & $  22.0892362690 $ & $   0.0000666806 $ & $   0.4035350405 $ & $   0.0119074602 $ & $   0.6190234684 $ & $   0.0046981111 $ & $   7.21 $ & \\
$f_{ 26 } $ & $  19.7878977340 $ & $   0.0000685054 $ & $   0.3796236632 $ & $   0.0119082300 $ & $   0.8120986841 $ & $   0.0049912433 $ & $   8.76 $ & $ f_{    2} +     2 f_{\rm orb} $ \\
$f_{ 27 } $ & $  16.1296946590 $ & $   0.0000705561 $ & $   0.3953352709 $ & $   0.0119826804 $ & $   0.4832366171 $ & $   0.0048238199 $ & $   8.99 $ & $ f_{    2}    -3 f_{\rm orb} $ \\
$f_{ 28 } $ & $  22.2773428300 $ & $   0.0000733798 $ & $   0.3603134859 $ & $   0.0119467153 $ & $   0.1635075211 $ & $   0.0052786054 $ & $   8.53 $ & $ f_{    1} +     7 f_{\rm orb} $ \\
$f_{ 29 } $ & $  12.0352796650 $ & $   0.0000757983 $ & $   0.3389022830 $ & $   0.0118968716 $ & $   0.6371364174 $ & $   0.0055874488 $ & $  10.88 $ & $ f_{    1}    -7 f_{\rm orb} $ \\
$f_{ 30 } $ & $  17.5506600900 $ & $   0.0000757425 $ & $   0.3318326737 $ & $   0.0149162795 $ & $   0.3713046176 $ & $   0.0071537202 $ & $   7.20 $ & $ f_{    1} +     6 f_{    7} $ \\
$f_{ 31 } $ & $  20.0442876680 $ & $   0.0000757988 $ & $   0.3235396648 $ & $   0.0120340866 $ & $   0.0370394995 $ & $   0.0059188433 $ & $   9.87 $ & \\
$f_{ 32 } $ & $  16.8817932390 $ & $   0.0000773410 $ & $   0.3214809049 $ & $   0.0126364832 $ & $   0.4604347814 $ & $   0.0062557301 $ & $   6.26 $ & $    3 f_{    7} +  f_{   19} $ \\
$f_{ 33 } $ & $  37.5583502060 $ & $   0.0000817451 $ & $   0.3152875146 $ & $   0.0119146155 $ & $   0.2122347001 $ & $   0.0060148187 $ & $   7.88 $ & \\
$f_{ 34 } $ & $  32.2504654220 $ & $   0.0000786696 $ & $   0.3123621267 $ & $   0.0121964216 $ & $   0.1494068796 $ & $   0.0062138766 $ & $   8.37 $ & $    2 f_{    4}    -2 f_{   32} $ \\
$f_{ 35 } $ & $  11.3036603290 $ & $   0.0000836472 $ & $   0.2963766096 $ & $   0.0118926509 $ & $   0.5441085363 $ & $   0.0063906508 $ & $   9.67 $ & $ f_{    1}    -8 f_{\rm orb} $ \\
$f_{ 36 } $ & $  36.4934262150 $ & $   0.0000858694 $ & $   0.3015737589 $ & $   0.0129140216 $ & $   0.2531474599 $ & $   0.0068138843 $ & $  18.15 $ & $    2 f_{    3}$ \\
$f_{ 37 } $ & $  16.8342557120 $ & $   0.0000849793 $ & $   0.3224052350 $ & $   0.0126091690 $ & $   0.1054265801 $ & $   0.0062231396 $ & $   5.98 $ & $    -5 f_{   19} +     5 f_{   31} $ \\
$f_{ 38 } $ & $  17.7466604430 $ & $   0.0000907109 $ & $   0.2752703351 $ & $   0.0119652926 $ & $   0.8976995791 $ & $   0.0069165264 $ & $   6.10 $ & $ f_{    1} +     9 f_{    7} $ \\
$f_{ 39 } $ & $  40.1499448370 $ & $   0.0000925404 $ & $   0.2836998368 $ & $   0.0127498545 $ & $   0.4900103612 $ & $   0.0071517113 $ & $   9.39 $ & $    2 f_{    3} +    5 f_{\rm orb} $ \\
$f_{ 40 } $ & $  16.7950430710 $ & $   0.0000988629 $ & $   0.2301770844 $ & $   0.0127566249 $ & $   0.3415268480 $ & $   0.0088222190 $ & $   6.45 $ & $ f_{   22}    -2 f_{\rm orb} $ \\
$f_{ 41 } $ & $  23.0144356590 $ & $   0.0000985161 $ & $   0.2192420954 $ & $   0.0122472133 $ & $   0.3102413705 $ & $   0.0088857219 $ & $   6.31 $ & $ f_{    1} +     8 f_{\rm orb} $ \\
$f_{ 42 } $ & $  37.4381658190 $ & $   0.0001001613 $ & $   0.2449822347 $ & $   0.0119153042 $ & $   0.9894855570 $ & $   0.0077405934 $ & $   7.46 $ & \\
$f_{ 43 } $ & $  23.1580380580 $ & $   0.0001040815 $ & $   0.2287372859 $ & $   0.0119550096 $ & $   0.8384059838 $ & $   0.0083142876 $ & $   6.76 $ & \\
$f_{ 44 } $ & $  21.3092764940 $ & $   0.0001047064 $ & $   0.2276616338 $ & $   0.0119138079 $ & $   0.6731238429 $ & $   0.0083252371 $ & $   6.74 $ & $ f_{   23}   -15 f_{\rm orb} $ \\
$f_{ 45 } $ & $  38.6868906280 $ & $   0.0001068949 $ & $   0.2251387641 $ & $   0.0118991205 $ & $   0.4567086326 $ & $   0.0084118533 $ & $   7.60 $ & $    2 f_{    3} +    3 f_{\rm orb} $ \\
$f_{ 46 } $ & $  41.9900799310 $ & $   0.0001081428 $ & $   0.2083561927 $ & $   0.0120638830 $ & $   0.9019115979 $ & $   0.0092152304 $ & $   9.36 $ & $    3 f_{   19}   -11 f_{\rm orb} $ \\
$f_{ 47 } $ & $  27.6354840030 $ & $   0.0001093502 $ & $   0.2160623230 $ & $   0.0119087175 $ & $   0.0874108749 $ & $   0.0087675021 $ & $   6.62 $ & $    7 f_{    3}    -6 f_{   19} $ \\
$f_{ 48 } $ & $  26.6070535980 $ & $   0.0001147005 $ & $   0.1942649420 $ & $   0.0119685671 $ & $   0.9752167764 $ & $   0.0098059184 $ & $   7.06 $ & $    -6 f_{    4} +     6 f_{   42} $ \\
$f_{ 49 } $ & $  29.5236278560 $ & $   0.0001141229 $ & $   0.2069591386 $ & $   0.0118984280 $ & $   0.7028060698 $ & $   0.0091493246 $ & $   6.92 $ & $    -2 f_{    3} +     2 f_{    4} $ \\
$f_{ 50 } $ & $  31.2222508790 $ & $   0.0001143168 $ & $   0.2067276436 $ & $   0.0119003068 $ & $   0.7342138109 $ & $   0.0091620040 $ & $   7.21 $ & $ f_{   24} +     4 f_{\rm orb} $ \\
$f_{ 51 } $ & $  18.2904917550 $ & $   0.0001114454 $ & $   0.2365868091 $ & $   0.0124162837 $ & $   0.6976326153 $ & $   0.0083531087 $ & $   5.12 $ & $   25 f_{\rm orb} $ \\
$f_{ 52 } $ & $   4.7345789640 $ & $   0.0001203252 $ & $   0.1937326504 $ & $   0.0118873767 $ & $   0.0425742064 $ & $   0.0097808112 $ & $   6.80 $ & \\
$f_{ 53 } $ & $  34.6127644020 $ & $   0.0001207062 $ & $   0.1686207872 $ & $   0.0122287226 $ & $   0.0427036284 $ & $   0.0115410075 $ & $   5.34 $ & $    2 f_{   43}   -16 f_{\rm orb} $ \\
$f_{ 54 } $ & $  16.8560934000 $ & $   0.0001168495 $ & $   0.2036739772 $ & $   0.0126521995 $ & $   0.9548025891 $ & $   0.0098897590 $ & $   5.28 $ & $ f_{    2}    -2 f_{\rm orb} $ \\
$f_{ 56 } $ & $  27.7463948590 $ & $   0.0001232307 $ & $   0.1893260537 $ & $   0.0119144159 $ & $   0.1086253138 $ & $   0.0100141092 $ & $   7.05 $ & $   -10 f_{    7} +     6 f_{   52} $ \\
$f_{ 57 } $ & $  16.5947524040 $ & $   0.0001235181 $ & $   0.1897377081 $ & $   0.0119445450 $ & $   0.9811652198 $ & $   0.0100080309 $ & $   5.89 $ & \\
$f_{ 58 } $ & $  17.5874722040 $ & $   0.0001244120 $ & $   0.1800091278 $ & $   0.0119950199 $ & $   0.0284799131 $ & $   0.0106098710 $ & $   6.00 $ & $ f_{    2} - f_{\rm orb} $ \\
$f_{ 59 } $ & $  18.1414042250 $ & $   0.0001216452 $ & $   0.1927523708 $ & $   0.0120056613 $ & $   0.1607895125 $ & $   0.0099151183 $ & $   5.20 $ & $ f_{   19} +     2 f_{\rm orb} $ \\
$f_{ 60 } $ & $  34.6997482550 $ & $   0.0001257662 $ & $   0.1728484313 $ & $   0.0120000845 $ & $   0.1754846720 $ & $   0.0110521779 $ & $   5.58 $ & \\
$f_{ 61 } $ & $  16.0946817180 $ & $   0.0001287149 $ & $   0.1786538420 $ & $   0.0119581100 $ & $   0.0797234333 $ & $   0.0106538450 $ & $   7.32 $ & $   22 f_{\rm orb} $ \\
$f_{ 62 } $ & $  18.3553034850 $ & $   0.0001301577 $ & $   0.1707763947 $ & $   0.0123495348 $ & $   0.0838395667 $ & $   0.0115130770 $ & $   5.26 $ & $ f_{   34}   -19 f_{\rm orb} $ \\
$f_{ 63 } $ & $  10.5720977270 $ & $   0.0001379949 $ & $   0.1660194106 $ & $   0.0119038319 $ & $   0.3512586753 $ & $   0.0114101739 $ & $   6.69 $ & $ f_{    1}    -9 f_{\rm orb} $ \\
        \hline
    \end{tabular}
\end{table*}

\begin{table*}
    \contcaption{}
    \label{tab:natbib}
    \begin{tabular}{lcccccccc}
        \hline
        ID & Frequency $f[\rm d^{-1}]$ & $\sigma_f [\rm d^{-1}]$ & Amplitude $A[\rm ppt]$ & $\sigma_A [\rm ppt]$ & Phase $\Phi [\rm rad]$ & $\sigma_{\Phi} [\rm rad]$ & S/N & Remarks \\
        \hline
        \hline
$f_{ 64 } $ & $  36.1004853800 $ & $   0.0001402758 $ & $   0.1631030226 $ & $   0.0119108837 $ & $   0.2393057587 $ & $   0.0116241415 $ & $   6.03 $ & $ f_{   33}    -2 f_{\rm orb} $ \\
$f_{ 65 } $ & $  29.0986104440 $ & $   0.0001407273 $ & $   0.1641385614 $ & $   0.0118956484 $ & $   0.4844784026 $ & $   0.0115379344 $ & $   6.52 $ & $ f_{   47} +     2 f_{\rm orb} $ \\
$f_{ 66 } $ & $  23.7408584650 $ & $   0.0001432008 $ & $   0.1583274819 $ & $   0.0118969116 $ & $   0.8398932133 $ & $   0.0119611014 $ & $   6.19 $ & $ f_{    1} +     9 f_{\rm orb} $ \\
$f_{ 67 } $ & $   9.8405715820 $ & $   0.0001466156 $ & $   0.1564808410 $ & $   0.0119012236 $ & $   0.1117887069 $ & $   0.0120988297 $ & $   6.60 $ & $ f_{    1}   -10 f_{\rm orb} $ \\
$f_{ 68 } $ & $  18.5813434970 $ & $   0.0001470171 $ & $   0.1538517471 $ & $   0.0119565937 $ & $   0.3343964329 $ & $   0.0123650486 $ & $   5.05 $ & $ f_{   31}    -2 f_{\rm orb} $ \\
$f_{ 69 } $ & $  25.1147424290 $ & $   0.0001554344 $ & $   0.1416063494 $ & $   0.0119071671 $ & $   0.4043785985 $ & $   0.0133810279 $ & $   5.30 $ & $ f_{   33}   -17 f_{\rm orb} $ \\
$f_{ 70 } $ & $  24.4724160770 $ & $   0.0001590826 $ & $   0.1430373159 $ & $   0.0118973921 $ & $   0.0982443302 $ & $   0.0132369983 $ & $   6.77 $ & $ f_{    1} +    10 f_{\rm orb} $ \\
$f_{ 71 } $ & $  32.0525513440 $ & $   0.0001614220 $ & $   0.1411480671 $ & $   0.0119226882 $ & $   0.5060906531 $ & $   0.0134467637 $ & $   4.82 $ & $ f_{   73}    -2 f_{\rm orb} $ \\
$f_{ 72 } $ & $  17.5126358240 $ & $   0.0001641864 $ & $   0.1542186029 $ & $   0.0122910076 $ & $   0.4357135866 $ & $   0.0126903703 $ & $   5.02 $ & $ f_{    3} - f_{\rm orb} $ \\
$f_{ 73 } $ & $  33.5155144560 $ & $   0.0001692705 $ & $   0.1418115231 $ & $   0.0119222954 $ & $   0.4694724266 $ & $   0.0133767490 $ & $   5.00 $ & $    5 f_{   7} +     2 f_{   57} $ \\
$f_{ 74 } $ & $  42.8624553010 $ & $   0.0001717528 $ & $   0.1293709576 $ & $   0.0118965066 $ & $   0.3405329993 $ & $   0.0146346820 $ & $   6.89 $ & $    2 f_{   19} +   13 f_{\rm orb} $ \\
$f_{ 75 } $ & $  35.4001193540 $ & $   0.0001714740 $ & $   0.1355692091 $ & $   0.0119087985 $ & $   0.1140401015 $ & $   0.0139852894 $ & $   4.89 $ & $    3 f_{   19}   -20 f_{\rm orb} $ \\
$f_{ 76 } $ & $  10.0355079930 $ & $   0.0001727101 $ & $   0.1313744163 $ & $   0.0119029275 $ & $   0.8360885315 $ & $   0.0144232879 $ & $   6.60 $ & $    8 f_{    7} +   13 f_{\rm orb} $ \\
$f_{ 78 } $ & $  22.9942826140 $ & $   0.0001714382 $ & $   0.1345406723 $ & $   0.0121982580 $ & $   0.5866739640 $ & $   0.0144299384 $ & $   4.64 $ & $ f_{   60}   -16 f_{\rm orb} $ \\
$f_{ 79 } $ & $  16.8129260140 $ & $   0.0001848728 $ & $   0.1410090954 $ & $   0.0126780212 $ & $   0.4957747396 $ & $   0.0143082932 $ & $   4.80 $ & $ f_{   30} - f_{\rm orb} $ \\
$f_{ 80 } $ & $  26.7856414350 $ & $   0.0001781572 $ & $   0.1195595817 $ & $   0.0119299981 $ & $   0.4694724266 $ & $   0.0158754678 $ & $   4.62 $ & $    -9 f_{   52} +     2 f_{   60} $ \\
$f_{ 81 } $ & $  27.2736456140 $ & $   0.0001776170 $ & $   0.1272237525 $ & $   0.0119146292 $ & $   0.4407382542 $ & $   0.0149054417 $ & $   5.03 $ & $    7 f_{    2}    -6 f_{   37} $ \\
$f_{ 82 } $ & $  33.7360560700 $ & $   0.0001790431 $ & $   0.1241458440 $ & $   0.0119186528 $ & $   0.8825306568 $ & $   0.0152797767 $ & $   4.81 $ & $ f_{    4} +  f_{\rm orb} $ \\
$f_{ 83 } $ & $  39.5830259160 $ & $   0.0001800485 $ & $   0.1215217588 $ & $   0.0119122766 $ & $   0.4663794539 $ & $   0.0156015961 $ & $   5.50 $ & $ f_{    4} +     9 f_{\rm orb} $ \\
$f_{ 84 } $ & $  35.1989665760 $ & $   0.0001807575 $ & $   0.1240083149 $ & $   0.0119057116 $ & $   0.6994131471 $ & $   0.0152826243 $ & $   5.12 $ & $ f_{    4} +     3 f_{\rm orb} $ \\
$f_{ 85 } $ & $  17.5436361850 $ & $   0.0002396458 $ & $   0.1724136340 $ & $   0.0152153703 $ & $   0.9538789111 $ & $   0.0140462599 $ & $   4.84 $ & $   -1 f_{    1} +  f_{   60} $ \\
$f_{ 86 } $ & $  37.5722710840 $ & $   0.0002190489 $ & $   0.1189898507 $ & $   0.0119175302 $ & $   0.0223133423 $ & $   0.0159387793 $ & $   4.44 $ & $    3 f_{    1}   -19 f_{\rm orb} $ \\
$f_{ 87 } $ & $  35.9305937150 $ & $   0.0001993510 $ & $   0.1148708052 $ & $   0.0119052653 $ & $   0.7919539214 $ & $   0.0164934570 $ & $   4.99 $ & $ f_{    4} +     4 f_{\rm orb} $ \\
$f_{ 88 } $ & $  31.5411274750 $ & $   0.0002014744 $ & $   0.1106051055 $ & $   0.0119025094 $ & $   0.6972763817 $ & $   0.0171291030 $ & $   5.42 $ & $ f_{    4}    -2 f_{\rm orb} $ \\
$f_{ 89 } $ & $  25.2038068670 $ & $   0.0002005821 $ & $   0.1123231770 $ & $   0.0119120601 $ & $   0.6366846021 $ & $   0.0168699102 $ & $   4.49 $ & $ f_{    1} +    11 f_{\rm orb} $ \\
$f_{ 90 } $ & $  22.2071276720 $ & $   0.0001746690 $ & $   0.1289386426 $ & $   0.0119517490 $ & $   0.6381933109 $ & $   0.0147456009 $ & $   4.00 $ & $ f_{   49}   -10 f_{\rm orb} $ \\
$f_{ 91 } $ & $  26.6612620160 $ & $   0.0002078542 $ & $   0.1108514153 $ & $   0.0119988954 $ & $   0.3815211671 $ & $   0.0172338096 $ & $   4.73 $ & $ f_{    1} +    13 f_{\rm orb} $ \\
$f_{ 92 } $ & $  42.0425978320 $ & $   0.0002044135 $ & $   0.1087721686 $ & $   0.0120633611 $ & $   0.6789248280 $ & $   0.0176516139 $ & $   5.67 $ & $    -6 f_{    3} +     9 f_{   37} $ \\
$f_{ 93 } $ & $  19.9725519320 $ & $   0.0002051852 $ & $   0.1077427077 $ & $   0.0120387231 $ & $   0.6155402717 $ & $   0.0177778918 $ & $   4.31 $ & $    7 f_{   52}   -18 f_{\rm orb} $ \\
$f_{ 94 } $ & $  38.5021663970 $ & $   0.0002094559 $ & $   0.1055528561 $ & $   0.0119002330 $ & $   0.6912824047 $ & $   0.0179423408 $ & $   4.50 $ & \\
$f_{ 95 } $ & $  39.9653470190 $ & $   0.0002116544 $ & $   0.1047246768 $ & $   0.0119184360 $ & $   0.9590103759 $ & $   0.0181102756 $ & $   4.72 $ & $ f_{   94} +     2 f_{\rm orb} $ \\
$f_{ 96 } $ & $  47.3833532690 $ & $   0.0002155664 $ & $   0.1036118874 $ & $   0.0118978268 $ & $   0.1350915658 $ & $   0.0182754872 $ & $   5.32 $ & $    3 f_{   38}    -8 f_{\rm orb} $ \\
$f_{ 97 } $ & $  43.9421078350 $ & $   0.0002173962 $ & $   0.1135353529 $ & $   0.0120471942 $ & $   0.1033860240 $ & $   0.0168901778 $ & $   5.06 $ & $ f_{    1} +  f_{   80} $ \\
$f_{ 98 } $ & $  34.6315019380 $ & $   0.0002266427 $ & $   0.1055100382 $ & $   0.0122268269 $ & $   0.5535460794 $ & $   0.0184411412 $ & $   4.08 $ & $ f_{   33}    -4 f_{\rm orb} $ \\
$f_{ 99 } $ & $   4.5691666540 $ & $   0.0002240292 $ & $   0.0985283523 $ & $   0.0119003143 $ & $   0.8365032224 $ & $   0.0192106527 $ & $   4.10 $ & $    9 f_{    2}    -8 f_{   31} $ \\
$f_{ 100 } $ & $  25.8276647670 $ & $   0.0002262715 $ & $   0.0971260444 $ & $   0.0118992093 $ & $   0.8116449481 $ & $   0.0195057774 $ & $   4.67 $ & $    7 f_{   52}   -10 f_{\rm orb} $ \\
$f_{ 102 } $ & $  30.0843337540 $ & $   0.0002352903 $ & $   0.0936406385 $ & $   0.0118989609 $ & $   0.7298839579 $ & $   0.0202210674 $ & $   4.15 $ & $ f_{   23}    -3 f_{\rm orb} $ \\
$f_{ 103 } $ & $  40.1657105750 $ & $   0.0002247442 $ & $   0.0994816840 $ & $   0.0127414425 $ & $   0.5739595477 $ & $   0.0203872102 $ & $   4.38 $ & $    2 f_{    1} +    8 f_{\rm orb} $ \\
$f_{ 104 } $ & $   8.3784430700 $ & $   0.0002543125 $ & $   0.0864111469 $ & $   0.0118959600 $ & $   0.8239244152 $ & $   0.0219106120 $ & $   4.67 $ & $ f_{    1}   -12 f_{\rm orb} $ \\
$f_{ 105 } $ & $  40.3072698180 $ & $   0.0002554254 $ & $   0.0867808774 $ & $   0.0119166611 $ & $   0.1850564655 $ & $   0.0218481819 $ & $   4.86 $ & $    2 f_{    2} +    5 f_{\rm orb} $ \\
$f_{ 106 } $ & $  16.0500502190 $ & $   0.0002610854 $ & $   0.0852547922 $ & $   0.0119321269 $ & $   0.6753112298 $ & $   0.0222717546 $ & $   4.06 $ & $ f_{    3}    -3 f_{\rm orb} $ \\
$f_{ 107 } $ & $  44.0082929640 $ & $   0.0002534593 $ & $   0.0868397072 $ & $   0.0120479371 $ & $   0.0152295717 $ & $   0.0220838150 $ & $   4.68 $ & $    2 f_{   32} +   14 f_{\rm orb} $ \\
$f_{ 108 } $ & $   9.1138769040 $ & $   0.0002673000 $ & $   0.0821270587 $ & $   0.0118956094 $ & $   0.0361225122 $ & $   0.0230545122 $ & $   4.37 $ & $ f_{    1}   -11 f_{\rm orb} $ \\
$f_{ 109 } $ & $  27.0374900740 $ & $   0.0002813125 $ & $   0.0780730321 $ & $   0.0119215453 $ & $   0.4250397675 $ & $   0.0242961248 $ & $   4.05 $ & $ f_{   22} +    12 f_{\rm orb} $ \\
$f_{ 110 } $ & $  46.9205107540 $ & $   0.0002966808 $ & $   0.0739975420 $ & $   0.0118980634 $ & $   0.5512875890 $ & $   0.0255884470 $ & $   4.06 $ & $    3 f_{    2}   -11 f_{\rm orb} $ \\
$f_{ 111 } $ & $  13.8883300960 $ & $   0.0003176941 $ & $   0.0707588978 $ & $   0.0119033593 $ & $   0.0952545136 $ & $   0.0267736564 $ & $   4.17 $ & $ f_{   30}    -5 f_{\rm orb} $ \\
$f_{ 112 } $ & $  13.2042298300 $ & $   0.0003057473 $ & $   0.0713939068 $ & $   0.0119073005 $ & $   0.8460216431 $ & $   0.0265429681 $ & $   4.08 $ & $ f_{    2}    -7 f_{\rm orb} $ \\
        \hline
    \end{tabular}
\end{table*}


\bsp    
\label{lastpage}
\end{document}